\documentclass[secthm,amsthm,pdftex]{elsarticle}

\usepackage{amsmath,amsfonts,amstext,amsthm,amssymb,latexsym}
\usepackage{enumitem,algorithm,algcompatible}
\usepackage{tikz}
\usepackage{tikz-qtree}

\newtheorem{theorem}{Theorem}[section]
\newtheorem{lemma}[theorem]{Lemma}

\theoremstyle{definition}
\newtheorem{definition}{Definition}[section]

\newtheorem{postulate}{Postulate}

\theoremstyle{remark}

\newcommand{\multiset}[1]{\{\!\!\{{#1}\}\!\!\}}

\journal{Science of Computer Programming}

\begin{document}
 
\begin{frontmatter}

\title{Behavioural Theory of Reflective Algorithms II: Reflective Parallel Algorithms\tnoteref{thanks}}

\author[1]{Klaus-Dieter Schewe\corref{a}\fnref{fn1}}
\ead{kdschewe@gmail.com}
\cortext[a]{Corresponding author}
\author[2]{Flavio Ferrarotti\fnref{fn2}}
\ead{flavio.ferrarotti@scch.at}

\fntext[fn1]{Part of the research of Klaus-Dieter Schewe has been supported by the ANR project EBRP:EventB-Rodin-Plus under grant no. ANR-19-CE25-0010.}
\fntext[fn2]{The research of Flavio Ferrarotti has been funded by the Federal Ministry for Climate Action, Environment, Energy, Mobility, Innovation and Technology (BMK), the Federal Ministry for Digital and Economic Affairs (BMDW), and the State of Upper Austria in the frame of the COMET Module Dependable Production Environments with Software Security (DEPS) within the COMET - Competence Centers for Excellent Technologies Programme managed by Austrian Research Promotion Agency FFG.}

\address[1]{Institut Nationale Polytechnique de Toulouse / IRIT CNRS, Toulouse, France}
\address[2]{Software Competence Centre Hagenberg, Hagenberg, Austria}

\begin{abstract}
We develop a behavioural theory of reflective parallel algorithms (RAs), i.e. synchronous parallel algorithms that can modify their own behaviour. The theory comprises a set of postulates defining the class of RAs, an abstract machine model, and the proof that all RAs are captured by this machine model. RAs are sequential-time, parallel algorithms, where every state includes a representation of the algorithm in that state, thus enabling linguistic reflection. Bounded exploration is preserved using multiset comprehension terms as values. The abstract machine model is defined by reflective Abstract State Machines (rASMs), which extend ASMs using extended states that include an updatable representation of the main ASM rule to be executed by the machine in that state.
\end{abstract}


\end{frontmatter}

\noindent
We dedicate this article to our former colleague, collaborator and friend Qing Wang (1972-2025). She will live on in our memories and our sincere appreciation for her personality and her inspiring research contributions.

\section{Introduction}

A {\em behavioural theory} in general comprises an axiomatic definition of a class of algorithms or systems by means of a set of characterising postulates, and an abstract machine model together with the proof that the abstract machine model captures the given class of algorithms or systems. The proof comprises two parts, one showing that every instance of the abstract machine model satisfies the postulates ({\em plausibility}), and another one showing that all algorithms stipulated by the postulates can be step-by-step simulated by an abstract machine, i.e. an instance of the abstract machine model ({\em characterisation}).

The ur-instance of a behavioural theory is Gurevich's sequential ASM thesis \cite{gurevich:tocl2000}, which clarifies what a sequential algorithm is and proves that sequential algorithms are captured by sequential Abstract State Machines (ASMs). Extensions of this thesis are the behavioural theories of parallel, recursive and concurrent algorithms \cite{ferrarotti:tcs2016,boerger:fi2020,boerger:ai2016} and further variations of these.

Adaptivity refers to the ability of a system to change its own behaviour. In the context of programming this concept is known as {\em (linguistic) reflection} since the 1950s. A brief survey of the development of reflection over the decades is contained in \cite{schewe:scp2021}. The recently increased interest in adaptive systems raises the question of their theoretical foundations. Such foundations are needed to achieve a common understanding of what can be gained by reflection, what the limitations of reflection are, and how reflection can be captured by state-based rigorous methods. This also constitutes the basis for the verification of properties of adaptive systems.

A first step towards answering this question was made by means of a behavioural theory of reflective sequential algorithms \cite{schewe:scp2022} and an extension of the logic of ASMs to a logic for reflective ASMs \cite{schewe:scp2021}. This theory provides the axiomatic definition of reflective sequential algorithms (RSAs), the proof that RSAs are not yet captured by Gurevich's postulates for sequential ASMs, the definition of reflective sequential ASMs (rsASMs), by means of which RSAs can be specified, and the proof that RSAs are captured by rsASMs, i.e. rsASMs satisfy the postulates of the axiomatisation, and any RSA stipulated by the axiomatisation can be defined by a behaviourally equivalent rsASM.

\subsection{Novel Contribution}

This article is dedicated to an extension of the theory to a behavioural theory of reflective, parallel algorithms (RAs), so the starting point is the simplified parallel ASM thesis\footnote{We prefer to exploit the simplified parallel ASM thesis rather than Blass's and Gurevich's parallel ASM thesis \cite{blass:tocl2003,blass:tocl2008}, because the axiomatisation is more compact, easier to comprehend and more convenient to use. Handling kens with mailboxes and displays, key concepts in the Blass/Gurevich axiomatisation, appears too cumbersome for the purpose of capturing reflection. Technically, both axiomatisations are equivalent, as both define the class of synchronous parallel algorithms, which are then captured by ASMs.} developed in \cite{ferrarotti:tcs2016}. Concerning the postulates it is crucial that abstract states of RAs must contain a representation of the algorithm itself, which can be interpreted as an executable rule that is to be used to define the successor state. This gives rise to generalised {\em sequential time} and {\em abstract state} postulates analogous to those for reflective sequential algorithms \cite{schewe:scp2022}.

Concerning a modified {\em bounded exploration} postulate we claim that while there cannot exist a fixed set of terms determining update sets in all states by simple interpretation, there is nonetheless a finite bounded exploration witness, provided a double interpretation is used: the first interpretation may result in multiset comprehension terms over the standard subsignature, which then can again be interpreted to define the values needed in the updates. Phrased differently, the first interpretation can be seen as resulting in a bounded exploration witness for the represented algorithm, and we obtain the terms that determine update sets by interpretation of generalised terms. Finally, these postulates involve some assumptions about the background, which are made explicit in a {\em background } postulate analogous to all other behavioural theories.

Concerning the definition of rASMs a concrete representation of a parallel ASM is required, for which we can choose a self-representation by means of trees. This allows us to exploit the tree algebra from Schewe and Wang \cite{schewe:jucs2010} to manipulate tree values as well as partial updates on tree values \cite{schewe:ejc2011}. The self-description comprising signature and rule can be stored in a dedicated location \textit{pgm\/}. Instead of defining update sets for a fixed rule, the rule to be considered is obtained by {\em raising} the value stored in the state representing the rule into an executable rule. 

For the plausibility and characterisation proofs the former one requires a construction of a bounded exploration witness from an rASM, for which the representation using  \textit{pgm\/} is essential, while the latter one will be accomplished by a sequence of lemmata, the key problem being that there is a theoretically unbounded number of different algorithms that nonetheless have to be handled uniformly. In essence we have to integrate the simplified parallel ASM thesis \cite{ferrarotti:tcs2016} and the reflective sequential ASM thesis \cite{schewe:scp2022}.

The work reported in this article extends a previous conference publication on the subject \cite{schewe:abz2025}, which does not contain the lengthy proofs. Proofs so far have been outsourced to a technical report \cite{schewe:arxiv2025} containing a preliminary version of this article.

\subsection{Organisation of this Article}

We start in Section \ref{sec:prelim} with some preliminaries concerning ASMs, the background of ASMs, and partial updates. For a detailed presentation of the concepts of ASMs we refer to B\"orger's and St\"ark's monograph \cite{boerger:2003}, while partial updates as we use them are handled in  \cite{schewe:ejc2011}. Section \ref{sec:ra} is dedicated to the first part of the behavioural theory, the axiomatic definition of RAs. The key problems concern the self-representation in abstract states and the extension of bounded exploration. In Section \ref{sec:trees} tree structures and a tree algebra are introduced following the work by Schewe and Wang \cite{schewe:jucs2010}. In Section \ref{sec:rasm} we introduce reflective ASMs, which are based on ASMs with the differences discussed above and a background structure capturing tree structures and tree algebra operations. We also show that rASMs satisfy the postulates from Section \ref{sec:ra}, thus they define RAs. Section \ref{sec:theory} is dedicated to the reflective parallel ASM thesis, i.e. with the more difficult part of the theory, the proof that every RA as stipulated by the postulates can be modelled by a behaviourally equivalent rASM. Section \ref{sec:others} contains a discussion of related work  focusing on linguistic reflection and other behavioural theories. We conclude with a summary and outlook in Section \ref{sec:schluss}.

\section{Preliminaries}\label{sec:prelim}

In this section we first recall some basic definitions about synchronous parallel ASMs \cite{boerger:2003}. We emphasise background structures and extend the rules of ASMs slightly by using partial updates \cite{schewe:ejc2011}, which are particularly useful when dealing with bulk structures such as trees.

\subsection{ASMs}

Fix a signature $\Sigma$, i.e. a finite set of function symbols, such that each $f \in \Sigma$ has an arity $ar(f)$. A {\em state} $S$ over $\Sigma$ is given by a base set $B$ and an interpretation of the function symbols in $\Sigma$ by functions $f_S : B^n \rightarrow B$ for $n=ar(f)$. The base set $B$ must contain truth values $0$ (false) and $1$ (true) as well as a special value \textit{undef\/}, which is used to capture partial functions.

An {\em isomorphism} between states $S$ and $S^\prime$ is a bijection $\pi: B \rightarrow B^\prime$ between the base sets of $S$ and $S^\prime$, respectively, such that $f_{S^\prime}(\pi(b_1), \dots, \pi(b_n)) = \pi(f_S(b_1, \dots, b_n))$ holds for all $f \in \Sigma$ and all $b_1, \dots, b_n \in B$. Naturally, we request that isomorphisms preserve truth values $0, 1$ and \textit{undef\/}.

Using $\Sigma$ and a set $X$ of variables we can define {\em terms} in the usual way. Then we obtain an evaluation function, which defines for each term $t \in \mathbb{T}$ and a variable assignement $\zeta: X \rightarrow B$ its value $\text{val}_{S,\zeta}(t)$ in a state $S$. Terms that are always interpreted by $0$, $1$ or \textit{undef\/} are referred to {\em Boolean terms}.

A sequential ASM {\em rule} over $\Sigma$ is defined as follows:

\begin{description}

\item[assignment.] Whenever $t_i$ ($i=0,\dots,n$) are terms over $\Sigma$ and $f \in \Sigma$ has arity $n$, then $f(t_1,\dots,t_n) := t_0$ is a rule.

\item[branching.] If $r_+$ and $r_-$ are rules and $\varphi$ is a Boolean term, then also \texttt{IF} $\varphi$ \texttt{THEN} $r_+$ \texttt{ELSE} $r_-$ \texttt{ENDIF} is a rule.

\item[bounded parallel composition.] If $r_1, \dots, r_k$ are rules, then also \texttt{PAR} $r_1 \dots r_k$ \texttt{ENDPAR} is a rule.

\item[parallel composition.] If $r(x)$ is a rule parameterised with a variable $x \in X$, and $\varphi(x)$ is a Boolean term with free variable $x$, then also \texttt{FORALL} $x$ \texttt{WITH} $\varphi(x)$ \texttt{DO} $r(x)$ \texttt{ENDDO} is a rule.

\item[let.] If $r(x)$ is a rule with free variable $x$ and $t$ a term, then \texttt{LET} $x = t$ \texttt{IN} $r(x)$ is a rule.

\item[import.] If $r(x)$ is a rule with free variable $x$, then \texttt{IMPORT} $x$ \texttt{DO} $r(x)$ is a rule.

\end{description}

In a paraterised rule $r(x)$ the variable $x$ is treated as a function symbol of arity $0$. The let rules are just a syntactic construct that eases definitions, but technically they can be omitted.

Each rule can be interpreted in a state, and doing so yields an update set. In general, a {\em location} is a pair $\ell = (f,(a_1,\dots,a_n))$ with a function symbol $f \in \Sigma$ of arity $n$ and an $n$-tuple of values from the base set $B$. An {\em update} is a pair $(\ell,a_0)$ with a value $a_0 \in B$. Given a variable assignment $\zeta: X \rightarrow B$ and a state $S$ with base set $B$ the update set $\Delta_{r,\zeta}(S)$ defined by a rule $r$ is yielded as follows:

\begin{itemize}

\item If $r$ is an assignment rule $f(t_1,\dots,t_n) := t_0$, then
\[ \Delta_{r,\zeta}(S) = \{ ((f,(\text{val}_{S,\zeta}(t_1) ,\dots, \text{val}_{rS\zeta}(t_n))), \text{val}_{S,\zeta}(t_0)) \} . \]

\item If $r$ is a branching rule \texttt{IF} $\varphi$ \texttt{THEN} $r_+$ \texttt{ELSE} $r_-$ \texttt{ENDIF}, then
\[ \Delta_{r,\zeta}(S) = \begin{cases} 
\Delta_{r_+,\zeta}(S) &\text{if}\; \text{val}_{S,\zeta}(\varphi) =1 \\
\Delta_{r_-,\zeta}(S) &\text{if}\; \text{val}_{S,\zeta}(\varphi) =0 \end{cases}
\]

\item If $r$ is a parallel rule \texttt{PAR} $r_1 \dots r_k$ \texttt{ENDPAR}, then $\Delta_{r,\zeta}(S) = \bigcup_{i=1}^k \Delta_{r_i,\zeta}(S)$.

\item If $r$ is a parallel rule  \texttt{FORALL} $x$ \texttt{WITH} $\varphi(x)$ \texttt{DO} $r^\prime(x)$ \texttt{ENDDO}, then $\Delta_{r,\zeta}(S) = \bigcup_{a \in \{ b \in B \mid \text{val}_{S,\zeta[x \mapsto b]}(\varphi(x)) = 1 \}} \Delta_{r(a),\zeta}(S)$.

\item If $r$ is a let rule \texttt{LET} $x = t$ \texttt{IN} $r^\prime$, then substituting $t$ for $x$ in $r^\prime$ defines $\Delta_{r,\zeta}(S) = \Delta_{\{ x \mapsto t \}.r^\prime,\zeta}(S)$.

\item If $r$ is an import rule \texttt{IMPORT} $x$ \texttt{DO} $r^\prime(x)$, then $\Delta_{r,\zeta}(S) = \Delta_{r^\prime(x),\zeta}(S)$, where $x$ is treated as a nullary function symbol that cannot appear as a location in an update set.

\end{itemize}

If $r$ is closed, i.e. it does not contain free variables, we can omit the variable assignment $\zeta$ and simply write $\Delta_r(S)$ instead of $\Delta_{r,\zeta}(S)$.

An update set $\Delta$ is {\em consistent} iff it does not contain clashes, i.e.\@ whenever $(\ell,v_1), (\ell,v_2) \in \Delta$ hold, then we must have $v_1 = v_2$. If $\Delta_r(S)$ is consistent, it defines a {\em successor state} $S^\prime = S + \Delta_r(S)$ by
\[ val_{S^\prime}(\ell) = \begin{cases} v_0 &\text{for}\; (\ell,v_0) \in \Delta_r(S) \\
val_S(\ell) &\text{else} \end{cases} \]

In accordance with \cite{boerger:2003} we extend this definition by $S + \Delta_r(S) = S$ in case $\Delta_r(S)$ is inconsistent.

\begin{definition}\rm

An {\em ASM} comprises a signature $\Sigma$ defining the set $\mathcal{S}$ of states, a subset $\mathcal{I} \subseteq \mathcal{S}$ of initial states over $\Sigma$, a closed ASM rule $r$ over $\Sigma$ and a transition function $\tau$ on states over $\Sigma$ with $\tau(S) = S + \Delta_r(S)$ for all states $S$. Both $\mathcal{S}$ and $\mathcal{I}$ are closed under isomorphisms.

\end{definition}

This definition of successor states gives rise to the notion of a run. A {\em run} of an ASM $M$ is a sequence $S_0, S_1, \dots$ of states with $S_0 \in \mathcal{I}$ and each $S_{i+1}$ being the successor of $S_i$.

\subsection{The Background}

The {\em background} of an ASM provides sets of values and operations on them that are part of every base set. For parallel ASMs we usually make some assumptions about their background without further mentioning them. First, not all values of the base set $B$ are used in a state $S$. We therefore assume a set {\em reserve} containing the unused values of the base set. We also mentioned already that in order to support partial functions we assume a value \textit{undef\/} in every base set. To avoid conflicts with non-strict functions we further assume that $f_S (v_1, \dots, v_n) = \textit{undef\/}$ holds, whenever one of the arguments $v_i$ is \textit{undef\/}. 

Furthermore, truth values 1 and 0 are also assumed to occur in every base set $B$, and we assume that the usual logical operators $\wedge$, $\neg$, $\vee$, etc. are defined. These as well as equality can be used in the same way as the function symbols in $\Sigma$ in the definition of terms. In this way we extend the set $\mathbb{T}(X)$ of terms and the evaluation function $\text{val}_{S,\zeta}$ in a state $S$ for a set $X$ of variables and a variable assignment $\zeta: X \rightarrow B$. 

In general, constants and operations of the background are are defined by background classes. According to \cite{blass:beatcs2007} a background class is determined by a background signature consisting of constructor and function symbols, the latter ones associated with a fixed arity, while for constructor symbols it is also permitted that the arity is unfixed or bounded.

\begin{definition}\label{def-bg-class}\rm

Let $\mathcal{D}$ be a set of base sets and $V_K$ a background signature, then a {\em background class} $\mathcal{K}$ with signature $V_K$ over $\mathcal{D}$ comprises a universe $U$ and an interpretation of function symbols in $V_K$ over $U$. The universe is defined as $U =
\bigcup \mathcal{D}^\prime$, where $\mathcal{D}^\prime$ is the smallest set with $\mathcal{D} \subseteq \mathcal{D}^\prime$ satisfying the following properties for each constructor symbol
$\llcorner\lrcorner\in V_K$:

\begin{itemize}

\item If $\llcorner\lrcorner\in V_K$ has unfixed arity, then $\llcorner D \lrcorner \in \mathcal{D}^\prime$ holds for all $D \in \mathcal{D}^\prime$ with
\[ {\llcorner D \lrcorner} = \{ \llcorner a_1,\dots,a_m \lrcorner \mid m \in \mathbb{N}, a_1,\dots,a_m \in D \} \]
and $A_{\llcorner \lrcorner} \in \mathcal{D}^\prime$ holds with $A_{\llcorner \lrcorner}=\bigcup_{\llcorner  D \lrcorner\in \mathcal{D}^\prime} \llcorner D \lrcorner$.

\item If $\llcorner\lrcorner\in V_K$ has bounded arity $n$, then $\llcorner D_1 ,\dots, D_m \lrcorner \in \mathcal{D}^\prime$ for all $m \le n$ and $D_i \in \mathcal{D}^\prime$ ($1 \le i \le m$) with
\[ {\llcorner D_1 ,\dots, D_m \lrcorner} = \{ \llcorner a_1,\dots,a_m \lrcorner \mid a_i \in D_i \;\text{for all}\; 1 \le i \le m \} \; . \]

\item If $\llcorner\lrcorner\in V_K$ has fixed arity $n$, then $\llcorner D_1 ,\dots, D_n \lrcorner \in \mathcal{D}^\prime$ for all $D_i \in \mathcal{D}^\prime$ with
\[ {\llcorner D_1 ,\dots, D_n \lrcorner} = \{ \llcorner a_1,\dots,a_n \lrcorner \mid a_i \in D_i \;\text{for all}\; 1 \le i \le n \} \; . \]

\end{itemize}

\end{definition}

The common {\em list} constructor $[\cdot]$ is an example of a constructor of unfixed arity; the same applies to a constructor $\langle\cdot\rangle$ for {\em finite multisets}. A constructor $(\cdot)_n$ for $n$-tuples is an example of a constructor with fixed arity $n$. Section \ref{sec:trees} is dedicated entirely to defining a background class of trees with a tree constructor and function symbols defining operations on trees.

The background of a parallel ASM must contain an infinite set \textit{reserve\/} of reserve values, truth values and their connectives, the equality predicate, the undefinedness value {\em undef}, and a background class $\mathcal{K}$ defined by a background signature $V_K$. In accordance with the parallel ASM theses \cite{blass:tocl2003,blass:tocl2008,ferrarotti:tcs2016} he background class must at least contain constructors for tuples and multisets and the appropriate operations on tuple and multiset values.

For convenience we further allow that constant symbols, i.e. nullary function symbols with the same interpretation in all states, are identified with the corresponding values in the base set\footnote{For instance, if we use natural numbers as constants, we do not make distinction between the value $5$ in the base set and a nullary function $5$ that is to interpreted by $5$.}, so they define ground terms that are interpreted by themselves. 

\subsection{Partial Updates}

The presence of a background class permits to have complex values such as sequences of arbitrary length, trees, graphs, etc. An update may only affect a tiny part of a complex value, but the presence of parallelism, though bounded, in sequential ASMs may cause avoidable conflicts on locations bound to such a complex value. For instance, if two updates in parallel affect only separate subtrees, they could be combined into a single tree update. We therefore add the following partial assignment rule to the definition of sequential ASM rules:

\begin{description}

\item[partial assignment.] Whenever $f \in \Sigma$ has arity $n$, $op \in V_K$ is an operator (i.e. a function symbol defined in the background) of arity $m+1$, $t_i$ ($i=1,\dots,n$) and $t_i^\prime$ ($i=1,\dots,m$) are terms over $\Sigma$, then $f(t_1,\dots,t_n) \leftleftarrows^{op} t_1^\prime ,\dots, t_m^\prime$ is a rule.

\end{description}

The effect of a single partial assignment can be captured by a single update
\[ ((f,(\text{val}_S(t_1) ,\dots, \text{val}_S(t_n))), \text{op}(\text{val}_S(f(t_1,\dots,t_n)),\text{val}_S(t_1^\prime) ,\dots, \text{val}_S(t_m^\prime))) \]

However, updates concerning the same location $\ell$ produced by partial assignments are first collected in a multiset $\ddot{\Delta}_\ell$. More precisely, the operators $op$ and their arguments $v_1^\prime ,\dots,v_m^\prime$ (with $v_i^\prime = \text{val}_S(t_i^\prime)$) will be collected in $\ddot{\Delta}_\ell$, i.e. $\ddot{\Delta}_\ell$ takes the form $\langle (\ell,op_1,(v_1^1 ,\dots,v_{m_1}^1)) , \dots , (\ell,op_k,(v_1^k ,\dots,v_{m_k}^k)) \rangle$. Any member of an update multiset $\ddot{\Delta}_\ell$ of the form $(\ell,op,(v_1 ,\dots,v_{m}))$ is called a {\em shared update} \cite{schewe:ejc2011}.

If possible, i.e. if the operators and arguments are compatible with each other, this multiset together with $\text{val}_S(f(t_1,\dots,t_n))$ will be collapsed into a single update $(\ell,v_0)$. Conditions for compatibility and the collapse of an update multiset into an update set have been elaborated in detail by Schewe and Wang \cite{schewe:ejc2011}. 

With the presence of partial update rules, the rule $r$ of an ASM $M$ first yields an update multiset $\ddot{\Delta}_r(S)$ for every state $S$ of $M$, which is defined analogously to update sets $\Delta_r(S)$ above. In addition to updates such an update multiset also contains shared updates. Collapsing all shared updates in $\ddot{\Delta}_r(S)$ for the same location, if possible, results in an update set $\Delta_r(S)$, which is uniquely determined by $\ddot{\Delta}_r(S)$.

Furthermore, with bulk values such as trees it is advisable to consider also fragments of values in connection with {\em sublocations}, which leads to dependencies among updates. For a tree $t$ every node $o \in \mathcal{O}$ defines such a sublocation, i.e. a nullary function symbol. In the Section \ref{sec:trees} we will make this explicit by providing functions that map values $o \in \mathcal{O}$ to such function symbols and vice-versa. Clearly, the value associated with the root determines the values associated with all sublocations. This is exploited in \cite{schewe:ejc2011} for the analysis of compatibility beyond the permutation of operators, but also on the level of sublocations. For this the notions of \emph{subsumption} and \emph{dependence} between locations are decisive.

\begin{definition}\label{def-subsumption}\rm

A location $\ell_1$ \emph{subsumes} a location $\ell_2$ (notation: $\ell_2 \sqsubseteq \ell_1$) iff for all states $S$ $val_S(\ell_1)$ uniquely determines $val_S(\ell_2)$. A location $\ell_1$ \emph{depends} on a location $\ell_2$ (notation: $\ell_2 \unlhd \ell_1$) iff $val_S(\ell_2) = \textit{undef\/}$ implies $val_S(\ell_1) = \textit{undef\/}$ for all states $S$.

\end{definition}

Clearly, for locations $\ell_1, \ell_2$ with $\ell_2 \sqsubseteq \ell_1$ we also have $\ell_1 \unlhd \ell_2$.

\section{Axiomatisation of Reflective Algorithms}\label{sec:ra}

We will define andmotivate the postulates for the class of reflective parallel algorithms. For this we will modify the postulates of synchronous parallel algorithms \cite{ferrarotti:tcs2016} to capture the requirements of linguistic reflection.

\subsection{Sequential Time}

Reflective algorithms proceed in sequential time, though in every step the behaviour of the algorithm may change. As argued in \cite{schewe:scp2022} it is always possible to have a finite representation of a sequential algorithm, which follows from the sequential ASM thesis in \cite{gurevich:tocl2000}. This does not change, if parallel algorithms as defined in \cite{ferrarotti:tcs2016} are considered, which is a consequence of any of the parallel ASM theses \cite{ferrarotti:tcs2016,blass:tocl2003,blass:tocl2008}, so the crucial feature of reflection can be subsumed in the notion of state, while the sequential time postulate can remain unchanged.

\begin{postulate}[Sequential Time Postulate]\label{p-time}\rm
A {\em reflective algorithm} (RA) comprises a set $\mathcal{S}$ of {\em states}, a subset $\mathcal{I} \subseteq \mathcal{S}$ of {\em initial states}, and a {\em one-step transition function} $\tau: \mathcal{S} \rightarrow \mathcal{S}$. Whenever $\tau(S) = S^\prime$ holds, the state $S^\prime$ is called the {\em successor state} of the state $S$.

\end{postulate}

A {\em run} of a reflective algorithm $\mathcal{A}$ is then given by a sequence $S_0, S_1, \dots$ of states $S_i \in \mathcal{S}$ with an initial state $S_0 \in \mathcal{I}$ and $S_{i+1} = \tau(S_i)$.

\subsection{Abstract States}

In order to capture reflection it will be necessary to modify the abstract state postulate such to capture the self-representation of the algorithm by a subsignature and to allow the signature to change. Furthermore, we need to be able to store terms as values, so the base sets need to be extended as well. For initial states we apply restrictions to ensure that the algorithm represented in an initial state is always the same. This gives rise to the following modification of the abstract state postulate.

\begin{postulate}[Abstract State Postulate]\label{p-state}\rm
States of an RA $\mathcal{A}$ must satisfy the following conditions:

\begin{enumerate}

\item Each state $S \in \mathcal{S}$ of $\mathcal{A}$ is a structure over some finite signature $\Sigma_S$, and an extended base set $B_{ext}$. The extended base set $B_{ext}$ contains at least a {\em standard base set} $B$ and all terms defined over $\Sigma_S$ and $B$.

\item The sets $\mathcal{S}$ and $\mathcal{I}$ of states and initial states of $\mathcal{A}$, respectively, are closed under isomorphisms. 

\item Whenever $\tau(S) = S^\prime$ holds, then $\Sigma_S \subseteq \Sigma_{\tau(S)}$, the states $S$ and $S^\prime$ of $\mathcal{A}$ have the same standard base set, and if $\sigma$ is an isomorphism defined on $S$, then also $\tau(\sigma(S)) = \sigma(\tau(S))$ holds.

\item For every state $S$ of $\mathcal{A}$ there exists a subsignature $\Sigma_{alg,S} \subseteq \Sigma_S$ for all $S$ and a function that maps the restriction of $S$ to $\Sigma_{alg,S}$ to a parallel algorithm $\mathcal{A}(S)$ with signature $\Sigma_S$, such that $\tau(S) = S + \Delta_{\mathcal{A}(S)}(S)$ holds for the successor state $\tau(S)$.

\item For all initial states $S_0, S_0^\prime \in \mathcal{I}$ we have $\mathcal{A}(S_0) = \mathcal{A}(S_0^\prime)$.

\end{enumerate}

\end{postulate}

Condition (iv) makes use of the unique minimal consistent update set $\Delta(S)$ defined by the algorithm in state $S$ satisfying $S + \Delta(S) = \tau(S)$. Concerning a state $S$ and its successor $\tau(S)$ we obtain a set $\textit{Diff\/}\, = \{ \ell \mid \text{val}_{\tau(S)}(\ell) \neq \text{val}_S(\ell) \}$ of those locations, where the states differ. Then $\Delta(S) = \{ (\ell,v) \mid \ell \in \textit{Diff\/} \wedge v = \text{val}_{\tau(S)}(\ell) \}$ is a consistent update set with $S + \Delta(S) = \tau(S)$ and $\Delta(S)$ is minimal with this property.

\subsection{Background}

We need to formulate minimum requirements for the {\em background}, which concern the reserve, truth values, tuples, as well as functions \textit{raise\/} and \textit{drop\/}, but they leave open how algorithms are represented by structures over $\Sigma_{alg,S}$. 

As backgrounds are defined by background structures, it is no problem to request that the set $\mathcal{D}$ of base sets contains an infinite set \textit{reserve\/} of reserve values and  the set $\mathbb{B} = \{ \mathbf{true}, \mathbf{false} \}$ of truth values. As we need to define arities for function symbols, $\mathcal{D}$ must further contain the set $\mathbb{N}$ of natural numbers. Note that truth values and natural numbers can be defined by hereditarily finite sets as in \cite{blass:apal1999}. Then the usual connectives on truth values need to be present as background function symbols. Furthermore, as in the parallel ASM thesis the background must contain constructors for tuples and multisets and the usual functions on them. As we leave the specific way of representing algorithms open, there may be further constructors and functions in the background, but they are not fixed.

We emphasised that we must be able to store terms as values, so instead of using an arbitrary base set $B$ we need an extended base set. For a state $S$ we denote by $B_{ext}$ the union of the universe $U$ defined by the background class $\mathcal{K}$ using $B$, $\mathbb{N}$, $\mathbb{B}$, and a subset of the reserve, and the set of all terms defined over $\Sigma_S$. We further denote by $\mathbb{T}_S$ the set of all terms defined over $\Sigma_S$. Note that with the presence of tuple and multiset constructors the set $\mathbb{T}_S$ contains also multiset comprehension terms as used in the parallel ASM thesis \cite{ferrarotti:tcs2016}. These are essential for obtaining bounded exploration witnesses.

In doing so we can treat a term in $\mathbb{T}_S$ as a term that can be evaluated in the state $S$ or simply as a value in $B_{ext}$. We use a function $\textit{drop\/}: \mathbb{T}_S \rightarrow B_{ext}$ that turns a term into a value of the extended base set, and a partial function $\textit{raise\/}: B_{ext} \rightarrow \mathbb{T}_S$ turning a value (representing a term) into a term that can be evaluated. In the same way we get a function $\textit{drop\/}: \mathcal{P}_S \rightarrow B_{ext}$ that turns an algorithm that can be executed in state $S$ into a value in the extended base set $B_{ext}$. Again, $\textit{raise\/}: B_{ext} \rightarrow \mathcal{P}_S$ denotes the (partial) inverse. We overload the function $\textit{drop\/}$ to also turn function symbols into values, i.e. we have $\textit{drop\/}: \Sigma_S \rightarrow B_{ext}$, for which $\textit{raise\/}: B_{ext} \rightarrow \Sigma_S$ denotes again the (partial) inverse. The presence of functions {\em raise} and {\em drop} is essential for linguistic reflection \cite{stemple:2000}.

\begin{postulate}[Background Postulate]\label{p-background}\rm
The {\em background of a RA} is defined by a background class $\mathcal{K}$ over a background signature $V_K$. It must contain an infinite set \textit{reserve\/} of reserve values, the equality predicate, the undefinedness value {\em undef}, truth values and their connectives, tuples and projection operations on them, multisets with union and comprehension operators on them, natural numbers and operations on them, and constructors and operators that permit the representation and update of parallel algorithms. 

The background must further provide partial functions: $\textit{drop\/}: \mathbb{T}_S \cup \mathcal{P}_S \cup \Sigma_S \rightarrow B_{ext}$ and $\textit{raise\/}: B_{ext} \rightarrow \mathbb{T}_S \cup \mathcal{P}_S \cup \Sigma_S$ for each base set $B$ and extended base set $B_{ext}$, and an {\em extraction function} $\beta: \mathbb{T}_S \rightarrow \bar{\mathbb{T}}$, which assigns to each term defined over a signature $\Sigma_S$ and the extended base set $B_{ext}$ a set of terms in $\bar{\mathbb{T}}$, which is defined over $B$, $\Sigma_S - \Sigma_{alg}$ and the tuple and multiset operators.

\end{postulate}

For instance, constructors for trees as well as operations on trees, e.g. for the extraction of subtrees or the composition of new trees must be defined in the background structures. We also need a set of constants such as \texttt{if}, \texttt{forall}, \texttt{par}, \texttt{let}, \texttt{assign} and \texttt{partial} by means of which we can label nodes in trees. Then expressions such as $tr = \langle \texttt{par}, \langle \texttt{assign}, c, \texttt{term} \langle f, t \rangle \rangle , t_2 \rangle$ are elements of $B_{ext}$ as well as executable algorithms in $\mathcal{P}_S$. While $tr$ itself is just a value, {\em raise}$(tr)$ is a rule that can be executed on the state $S$. In this case $\beta(tr)$ is the set of multiset terms defined by the terms $c, f(t), t$ occurring in the value $tr$.

\subsection{Bounded Exploration}

Finally, we need a generalisation of the bounded exploration postulate. As in every state $S$ we have a representation of the actual parallel algorithm $\mathcal{A}(S)$, this algorithm possesses a bounded exploration witness $W_S$, i.e. a finite set of multiset comprehension terms in $\bar{\mathbb{T}}$ such that $\ddot{\Delta}_{\mathcal{A}(S)}(S_1) = \ddot{\Delta}_{\mathcal{A}(S)}(S_2)$ (and thus also $\Delta_{\mathcal{A}(S)}(S_1) = \Delta_{\mathcal{A}(S)}(S_2)$) holds, whenever states $S_1$ and $S_2$ coincide on $W_S$. We can always assume that $W_S$ just contains terms that must be evaluated in a state to determine the update set in that state. Thus, though $W_S$ is not unique we may assume that $W_S$ is contained in the finite representation of $\mathcal{A}(S)$. This implies that the terms in $W_S$ result by interpretation from terms that appear in this representation, i.e. $W_S$ can be obtained using the extraction function $\beta$. Consequently, there must exist a finite set of terms $W$ such that its interpretation in a state yields both values and terms, and the latter ones represent $W_S$. We will continue to call $W$ a {\em bounded exploration witness}. Then the interpretation of $W$ and the interpretation of the extracted terms in any state suffice to determine the update set in that state. This leads to our {\em bounded exploration postulate} for RAs. 

If $S$, $S^\prime$ are states of an RA and $W$ is a set of multiset comprehension terms over the common signature $\Sigma_S \cap \Sigma_{S^\prime}$, we say that $S$ and $S^\prime$ {\em strongly coincide} over $W$ iff the following holds:

\begin{itemize}

\item For every $t \in W$ we have $\text{val}_S(t) = \text{val}_{S^\prime}(t)$.   

\item For every $t \in W$ with $\text{val}_S(t) \in \mathbb{T}_S$ and $\text{val}_{S^\prime}(t) \in \mathbb{T}_{S^\prime}$ we have \\
$\text{val}_S(\beta(t)) = \text{val}_{S^\prime}(\beta(t))$.
 
\end{itemize}

We may further assume that the complex values representing an algorithm are updated by several operations in one step, i.e. shared updates defined by an operator and arguments may be used to define updates \cite{schewe:ejc2011}. If operators are compatible, such shared updates are merged into a single update. In order to capture this merging we extend the bounded exploration witness $W$ as follows: A term indicating a shared update takes the form $op(f(t_1,\dots,t_n),t_1^\prime,\dots,t_m^\prime)$, where $op$ is the operator that is to be applied, $f(t_1,\dots,t_n)$ evaluates in every state $S$ to a value $\text{val}_S(\ell)$ of some location $\ell = (f, (\text{val}_S(t_1) ,\dots, \text{val}_S(t_n)))$, and $t_1^\prime,\dots,t_m^\prime$ evaluate to the other arguments of the shared update. 

If $op_1(f(t_1,\dots,t_n),t_{11}^\prime,\dots,t_{1m_1}^\prime) ,\dots, op_k(f(t_1,\dots,t_n),t_{k1}^\prime,\dots,$ $t_{km_k}^\prime)$ are several terms occurring in $W$ or in $\beta(W)$, then the term
\[ op_1(\dots (op_k(f(t_1,\dots,t_n),t_{k1}^\prime,\dots,t_{km_k}^\prime),\dots, 
t_{21}^\prime,\dots,t_{2m_2}^\prime), t_{11}^\prime,\dots,t_{1m_1}^\prime) \]
will be called an {\em aggregation term over $f(t_1,\dots,t_n)$}, and the tuple $(\text{val}_S(\hat{t})$, $ \text{val}_S(t_1) ,\dots, \text{val}_S(t_n))$ will be called an {\em aggregation tuple}. 
Then we can always assume that the update set $\Delta_{\mathcal{A}}(S)$ is the result of collapsing an update multiset $\ddot{\Delta}_{\mathcal{A}}(S)$.

\begin{postulate}[Bounded Exploration Postulate]\label{p-bounded}\rm
For every RA $\mathcal{A}$ there is a finite set $W$ of multiset comprehension terms such that $\ddot{\Delta}_{\mathcal{A}}(S) = \ddot{\Delta}_{\mathcal{A}}(S^\prime)$ holds (hence also $\Delta_{\mathcal{A}}(S) = \Delta_{\mathcal{A}}(S^\prime)$) whenever the states $S$ and $S^\prime$ strongly coincide over $W$.

Furthermore, $\ddot{\Delta}_{\mathcal{A}}(\text{res}(S,\Sigma_{alg})) = \ddot{\Delta}_{\mathcal{A}}(\text{res}(S^\prime,\Sigma_{alg}))$ holds (and consequently also $\Delta_{\mathcal{A}}(\text{res}(S,\Sigma_{alg})) = \Delta_{\mathcal{A}}(\text{res}(S^\prime,\Sigma_{alg}))$) whenever the states $S$ and $S^\prime$ coincide over $W$. Here $\text{res}(S,\Sigma_{alg})$ is the structure resulting from $S$ by restriction of the signature to $\Sigma_{alg}$.

\end{postulate}

Any set $W$ of terms as in the bounded exploration postulate \ref{p-bounded} will be called a {\em (reflective) bounded exploration witness} (R-witness) for $\mathcal{A}$. The four postulates capturing sequential time, abstract states, background and bounded exploration together provide an axiomatic definition of the notion of a reflective algorithm.

\subsection{Behavioural Equivalence}

According to the definitions in \cite{gurevich:tocl2000,blass:tocl2003} two algorithms are {\em behaviourally equivalent} iff they have the same sets of states and initial states and the same transition function $\tau$.  If we adopted without change this definition of behavioural equivalence from \cite{gurevich:tocl2000}, then the substructure over $\Sigma_{alg}$ would be required to be exactly the same in corresponding states. However, the way how to realise such a representation was deliberately left open in the axiomatisation. Therefore, instead of claiming identical states it suffices to only require identity for the restriction to $\Sigma_S - \Sigma_{alg}$, while structures over $\Sigma_{alg}$ only need to define behaviourally equivalent algorithms. 

However, this is still too restrictive, as behavioural equivalence of $\mathcal{A}(S)$ and $\mathcal{A}(S^\prime)$ would still imply identical changes to the self-representation. Therefore, we can also restrict our attention to the behaviour of the algorithms $\mathcal{A}(S)$ on states over $\Sigma_S - \Sigma_{alg}$. If $\mathcal{A}(S)$ and $\mathcal{A}(S^\prime)$ produce significantly different changes to the represented algorithm, the next state in a run of the reflective algorithm will reveal this. 

Therefore, two RAs $\mathcal{A}$ and $\mathcal{A}^\prime$ are {\em behaviourally equivalent} iff there exists a bijection $\Phi$ between runs of $\mathcal{A}$ and those of $\mathcal{A}^\prime$ such that for every run $S_0, S_1, \dots$ of $\mathcal{A}$ we have that for all states $S_i$ and $\Phi(S_i)$ 

\begin{enumerate}

\item their restrictions to $\Sigma_{S_i} - \Sigma_{alg}$ are the same, and

\item the restrictions of $\mathcal{A}(S_i)$ and $\mathcal{A}^\prime(\Phi(S_i))$ to $\Sigma_{S_i} - \Sigma_{alg}$ represent behaviourally equivalent parallel algorithms.

\end{enumerate}

\section{Tree Structures and Tree Algebra}\label{sec:trees}

In order to develop an abstract machine model capturing RSAs we introduce first tree structures and a tree algebra. These structures will be defined as part of the background of an rsASM in the next section.

\subsection{Tree Structures}

We now provide the details of the tree structures and the tree algebra. 

\begin{definition} \label{def-unranked-tree}

An {\em unranked tree structure} is a structure $(\mathcal{O},\prec_{c},\prec_{s})$ consisting of a finite, non-empty set $\mathcal{O}$ of node identifiers, called tree domain, and irreflexive relations $\prec_{c}$ (child relation) and $\prec_{s}$ (sibling relation) over $\mathcal{O}$ satisfying the following conditions:

\begin{itemize}

\item there exists a unique, distinguished node $o_r \in \mathcal{O}$ (root) such that for all $o \in \mathcal{O} - \{ o_r \}$ there is exactly one $o^\prime \in \mathcal{O}$ with $o^\prime \prec_c o$, and

\item whenever $o_1 \prec_s o_2$ holds, then there is some $o \in \mathcal{O}$ with $o \prec_c o_i$ for $i=1,2$.

\end{itemize}

\end{definition}

For $x_1 \prec_{c} x_2$ we say that $x_2$ is a {\em child} of $x_1$. For $x_1 \prec_{s} x_2$ we say that $x_2$ is the {\em next sibling} of $x_1$, and $x_1$ is the {\em previous sibling} of $x_2$. In order to obtain trees from this, we add labels and assign values to the leaves. For this we fix a finite, non-empty set $L$ of {\em labels}, and a finite family $\{ \tau_i \}_{i \in I}$ of data types. Each data type $\tau_i$ is associated with a {\em value domain} $dom(\tau_i)$. The corresponding {\em universe} $U$ contains all possible values of these data types, i.e. $U=\bigcup_{i \in I}dom(\tau_i)$.

\begin{definition}\label{def-tree}

A {\em tree} $t$ over the set of labels $L$ with values in the universe $U$ comprises an unranked tree structure $\gamma_t=(\mathcal{O}_t,\prec_{c},\prec_{s})$, a total label function $\omega_t: \mathcal{O}_t \rightarrow L$, and a partial value function $\upsilon_t: \mathcal{O}_t \rightarrow U$ that is defined on the leaves in $\gamma_t$.

\end{definition}

Let $T_L$ denote the set of all trees with labels in $L$, and let $root(t)$ denote the root node of a tree $t$. A sequence $t_1,...,t_k$ of trees is called a {\em hedge}, and a multiset $\langle t_1,...,t_k \rangle$ of trees is called a {\em forest}. Let $\epsilon$ denote the empty hedge, and let $H_L$ denote the set of all hedges with labels in $L$. A tree $t_1$ is a {\em subtree} of $t_2$ (notation $t_1 \sqsubseteq t_2$) iff the following properties are satisfied\footnote{Note that this definition of subtree allows some siblings to be omitted, which is more general than considering just the trees rooted at some node.}:

\begin{enumerate}

\item $\mathcal{O}_{t_1} \subseteq \mathcal{O}_{t_2}$, 

\item $o_1 \prec_c o_2$ holds in $t_1$ for $o_1, o_2 \in \mathcal{O}_{t_1}$ iff it holds in $t_2$, 

\item $o_1 \prec_s o_2$ holds in $t_1$ for $o_1, o_2 \in \mathcal{O}_{t_1}$ iff it holds in $t_2$, 

\item $\omega_{t_1}(o^\prime) = \omega_{t_2}(o^\prime)$ holds for all $o^\prime \in \mathcal{O}_{t_1}$, and 

\item for all leaves $o^\prime \in \mathcal{O}_{t_1}$ we have $\upsilon_{t_1}(o^\prime) = \upsilon_{t_2}(o^\prime)$. 

\end{enumerate}

$t_1$ is the {\em largest subtree} of $t_2$ (denoted as $\widehat{o}$) at node $o$ iff $t_1 \sqsubseteq t_2$ with $root(t_1) = o$ and there is no tree $t_3$ with $t_1 \neq t_3 \neq t_2$ such that $t_1 \sqsubseteq t_3 \sqsubseteq t_2$.  

\begin{definition}\label{def-context}

The {\em set of contexts} $C_L$ over $L$ is the set $T_{L \cup \{\xi\}}$ of trees with labels in $L \cup \{\xi\}$ ($\xi \notin L$) such that for each tree $t \in C_L$ exactly one leaf node is labelled with $\xi$ and the value assigned to this leaf is \textit{undef\/}.

\end{definition}

The context with a single node labelled $\xi$ is called trivial and is simply denoted as $\xi$. Contexts allow us to define substitution operations that replace a subtree of a tree or context by a new tree or context. This leads to the following four kinds of substitutions:

\begin{description}

\item[Tree-to-tree substitution.] For a tree $t_1 \in T_{L_1}$, a node $o \in \mathcal{O}_{t_1}$ and a tree $t_2 \in T_{L_2}$ the result $\textit{subst}_{tt}(t_1, o, t_2) = t_1[\widehat{o} \mapsto t_2]$ of substituting $t_2$ for the subtree rooted at $o$ is a tree in $T_{L_1 \cup L_2}$.

\item[Tree-to-context substitution.] For a tree $t_1 \in T_{L_1}$, a node $o \in \mathcal{O}_{t_1}$ the result $\textit{subst}_{tc}(t_1, o, \xi) = t_1[\widehat{o} \mapsto \xi]$ of substituting the trivial context for the subtree rooted at $o$ is a context in $C_{L_1}$.

\item[Context-to-context substitution.] For contexts $c_1 \in C_{L_1}$ and $c_2 \in C_{L_2}$ the result $\textit{subst}_{cc}(c_1, c_2) = c_1[\xi \mapsto c_2]$ of substituting $c_2$ for the leaf labelled by $\xi$ in $c_1$ is a context in $C_{L_1 \cup L_2}$.

\item[Context-to-tree substitution.] For a context $c_1 \in C_{L_1}$ and a tree $t_2 \in T_{L_2}$ the result $\textit{subst}_{ct}(c_1, t_2) = c_1[\xi \mapsto t_2]$ of substituting $t_2$ for the leaf labelled by $\xi$ in $c_1$ is a tree in $T_{L_1 \cup L_2}$.

\end{description}

As a shortcut we also write $\textit{subst}_{tc}(t_1, o, c_2)$ for $\textit{subst}_{cc}(\textit{subst}_{tc}(t_1, o, \xi), c_2)$, which is a context in $C_{L_1 \cup L_2}$.

\subsection{Tree Algebra}

To provide manipulation operations over trees at a level higher than individual nodes and edges, we need constructs to select arbitrary tree portions. For this we provide two selector constructs, which result in subtrees and contexts, respectively. For a tree $t = (\gamma_t, \omega_t, \upsilon_t) \in T_L$ these constructs are defined as follows:

\begin{itemize}

\item $context: \mathcal{O}_t \times \mathcal{O}_t \rightarrow C_L$ is a partial function on pairs $(o_1,o_2)$ of nodes with $o_1 \prec_c^+ o_2$ and $context(o_1,o_2) = \textit{subst\/}_{tc}(\widehat{o}_1, o_2, \xi) = \widehat{o}_1[{\widehat{o}_2\mapsto\xi}]$, where $\prec_c^+$ denotes the transitive closure of $\prec_c$.

\item $subtree: \mathcal{O}_t \rightarrow T_L$ is a function defined by $subtree(o) = \widehat{o}$.

\end{itemize}

The {\em set $\mathbb{T}$ of tree algebra terms} over $L \cup \{ \epsilon, \xi \}$ comprises label terms, hedge terms, and context terms, i.e. $\mathbb{T} = L \cup \mathbb{T}_h \cup \mathbb{T}_c$, which are defined as follows:

\begin{itemize}

\item The set $\mathbb{T}_h$ is the smallest set with $T_L \subseteq \mathbb{T}_h$ such that

\begin{enumerate}

\item $\epsilon \in \mathbb{T}_h$, 

\item $a \langle h \rangle \in \mathbb{T}_h$ for $a \in L$ and $h \in
\mathbb{T}_h$, and

\item $t_1 \dots t_n \in \mathbb{T}_h$ for $t_i \in \mathbb{T}_h$ ($i=1 ,\dots, n$).

\end{enumerate}

\item The set of context terms $\mathbb{T}_c$ is the smallest set with

\begin{enumerate}

\item $\xi \in \mathbb{T}_c$ and

\item $a \langle t_1 ,\dots, t_n \rangle \in \mathbb{T}_c$ for $a \in L$ and terms $t_1 ,\dots, t_n \in \mathbb{T}_h \cup \mathbb{T}_c$, such that exactly one $t_i$ is a context term in $\mathbb{T}_c$.

\end{enumerate}

\end{itemize}

With these we now define the operators of our tree algebra as follows:

\begin{description}

\item[label\_hedge.] The operator \textit{label\_hedge\/} turns a hedge into a tree with a new added root, i.e. $\textit{label\_hedge\/}(a, t_1 \dots t_n) = a \langle t_1, \dots, t_n \rangle$.

\item[label\_context.] Similarly, the operator \textit{label\_context\/} turns a context into a context with a new added root, i.e. $\textit{label\_context\/}(a, c) = a \langle c \rangle$.

\item[left\_extend.] The operator \textit{left\_extend\/} integrates the trees in a hedge into a context extending it on the left, i.e. \\
$\textit{left\_extend\/}(t_1 \dots t_n, a \langle t_1^\prime ,\dots, t_m^\prime \rangle) = a \langle t_1, \dots, t_n, t_1^\prime, \dots, t_m^\prime \rangle$.

\item[right\_extend.] Likewise, the operator \textit{right\_extend\/} integrates the trees in a hedge into a context extending it on the right, i.e.\\
$\textit{right\_extend\/}(t_1 \dots t_n, a \langle t_1^\prime ,\dots, t_m^\prime \rangle) = a \langle t_1^\prime, \dots, t_m^\prime, t_1, \dots, t_n \rangle$.

\item[concat.] The operator \textit{concat\/} simply concatenates two hedges, i.e. \\
$\textit{concat\/}(t_1 \dots t_n, t_1^\prime \dots t_m^\prime) = t_1 \dots t_n t_1^\prime \dots t_m^\prime$.

\item[inject\_hedge.] The operator \textit{inject\_hedge\/} turns a context into a tree by substituting a hedge for $\xi$, i.e. $\textit{inject\_hedge\/}(c, t_1 \dots t_n) = c[\xi \mapsto t_1 \dots t_n]$.

\item[inject\_context.] The operator \textit{inject\_context\/} substitutes a context for $\xi$, i.e. $\textit{inject\_context\/}(c_1,c_2) = c_1[\xi \mapsto c_2]$.

\end{description}

\section{Reflective Abstract State Machines}\label{sec:rasm}

In this section we define a model of reflective ASMs, which uses a self representation of an ASM as a particular tree value that is assigned to a location \textit{pgm\/}. For basic operations on this tree we exploit the tree algebra from Section \ref{sec:trees}. Then in every step the update set will be built using the rule in this representation, for which we exploit \textit{raise\/} and \textit{drop\/} as in \cite{stemple:2000}.

\subsection{Self Representation Using Trees}\label{ssec:treerep}

For the dedicated location storing the self-representation of an ASM it is sufficient to use a single function symbol \textit{pgm\/} of arity $0$. Then in every state $S$ the value $\text{val}_S(\textit{pgm\/})$ is a complex tree comprising two subtrees for the representation of the signature and the rule, respectively. The signature is just a list of function symbols, each having a name and an arity. The rule can be represented by a syntax tree. 

Thus, for the tree structure we have a root node $o$ labelled by \texttt{pgm} with exactly two successor nodes, say $o_0$ and $o_1$, labelled by \texttt{signature} and \texttt{rule}, respectively. So we have $o \prec_c o_0$, $o_0 \prec_s o_1$ and $o \prec_c o_1$. 

The subtree rooted at $o_0$ has as many children $o_{00} ,\dots, o_{0k}$ as there are function symbols in the signature, each labelled by \texttt{func}. Each of the subtrees rooted at $o_{0i}$ takes the form $\texttt{func} \langle \texttt{name} \langle f \rangle \; \texttt{arity} \langle n \rangle \rangle$ with a function name $f$ and a natural number $n$. 

The subtree rooted at $o_1$ represents the rule as a tree. Trees representing rules are inductively defined as follows:

\begin{itemize}

\item An assignment rule $f(t_1,\dots,t_n) = t_0$ is represented by a tree of the form $\textit{label\_hedge}(\texttt{update},\texttt{func} \langle f \rangle \texttt{term} \langle t_1 \dots t_n \rangle \texttt{term} \langle t_0 \rangle)$.

\item A partial assignment rule $f(t_1,\dots,t_n) \leftleftarrows^{op} t_1^\prime, \dots, t_m^\prime$ is represented by a tree of the form\\
$\textit{label\_hedge}(\texttt{partial},\texttt{func} \langle f \rangle \texttt{func} \langle op \rangle \texttt{term} \langle t_1 \dots t_n \rangle \texttt{term} \langle t_1^\prime \dots t_m^\prime \rangle)$.

\item A branching rule \texttt{IF} $\varphi$ \texttt{THEN} $r_1$ \texttt{ELSE} $r_2$ \texttt{ENDIF} is represented by a tree of the form $\textit{label\_hedge}(\texttt{if},\texttt{bool} \langle \varphi \rangle \texttt{rule} \langle t_1 \rangle \texttt{rule} \langle t_2 \rangle)$, where $t_i$ (for $i=1,2$) is the tree representing the rule $r_i$.

\item A bounded parallel rule \texttt{PAR} $r_1 \dots r_k$ \texttt{ENDPAR} is represented by a tree of the form $\textit{label\_hedge}(\texttt{par}, \texttt{rule} \langle t_1 \rangle \; \dots \; \texttt{rule} \langle t_k \rangle)$, where $t_i$ (for $i=1,\dots,k$) is the tree representing the rule $r_i$.

\item A parallel rule \texttt{FORALL} $x$ \texttt{WITH} $\varphi$ \texttt{DO} $r$ \texttt{ENDDO} is represented by a tree of the form $\textit{label\_hedge}(\texttt{forall}, \texttt{term}\langle x \rangle, \texttt{bool}\langle\varphi\rangle,
\texttt{rule}\langle t \rangle)$, where $t$ is the tree representing the rule $r$.

\item A let rule \texttt{LET} $x = t$ \texttt{IN} $r$ is represented by a tree of the form \\
$\textit{label\_hedge}(\texttt{let},\texttt{term} \langle x \rangle \texttt{term} \langle t \rangle \texttt{rule} \langle t^\prime \rangle)$, where $t^\prime$ is the tree representing the rule $r$.

\item An import rule \texttt{IMPORT} $x$ \texttt{DO} $r$ (which imports a fresh element from the reserve) is represented by a tree of the form $\textit{label\_hedge}(\texttt{import},\texttt{term} \langle x \rangle$ $\texttt{rule} \langle t \rangle)$, where $t$ is the tree representing the rule $r$. 

\end{itemize}

\subsection{The Background of an rASM}

Let us draw some consequences from this tree representation. As function names in the signature appear in the tree representation, these are values. Furthermore, we may in every step enlarge the signature, so there must be an infinite reserve of such function names. Since ASMs have an infinite reserve of values, new function names can  naturally be imported from this reserve. Likewise we require natural numbers in the background for the arity assigned to function symbols, though operations on natural numbers are optional. Terms built over the signature and a base set $B$ must also become values. This includes tuple and multiset terms. Concerning the subtree capturing the rule, the keywords for the different rules become labels, so we obtain the set of labels
\begin{align*}
L = &\{ \texttt{pgm},  \texttt{signature}, \texttt{rule}, \texttt{func}, \texttt{name}, \texttt{arity}, \texttt{update}, \texttt{term}, \texttt{if},\\
&\qquad \texttt{bool}, \texttt{forall}, \texttt{par}, \texttt{let}, \texttt{partial}, \texttt{import} \} .
\end{align*}

Consequently, we must extend a base set $B$ by such terms, i.e. terms will become values. For this let $\Sigma$ be a signature and let $B$ be the base set of a structure of signature $\Sigma$. An {\em extended base set} is the smallest set $B_{\textit{ext\/}}$ that includes all elements of $B$, all elements in the reserve, all natural numbers, all terms $t_i$ of signature $\Sigma \cup \Sigma _{\textit{ext\/}}$ such that $\Sigma_{\textit{ext\/}}$ is some finite subset of $\{raise(v_i) \mid v_i \text{ is an element of the reserve}\}$ and each function symbol of  $\Sigma_{\textit{ext\/}}$ has some arbitrary fixed arity, and all terms of the tree algebra over all possible signatures $\Sigma_{\textit{ext\/}}$ with labels in $L$ as defined above. As $B$ is closed under the operators in the background, also tuples and multisets are included in $B$ and hence in $B_{\textit{ext\/}}$.

In an extended base set terms are treated as values and thus can appear as values of some locations in a state. This implies that terms now have a dual character. When appearing in an ASM rule, e.g. on the right-hand side of an assignment, they are interpreted in the current state to determine an update. However, if they are to be treated as a value, they have to be interpreted by themselves. Therefore, we require a function \textit{drop\/} turning a term into a value, and inversely a function \textit{raise\/} turning a value into a term. Over elements of the (non-extended) base set $B$, elements imported from the reserve and natural numbers, i.e. over elements that do not represent terms, both functions can be thought of as the identity. Thus, if we raise an element $a$ of $B$, then the result is a nullary function symbol named $a$. Such a function symbol $a$ can be seen as a constant in the sense that it is always interpreted by itself, i.e. by $a \in B$. As discussed in~\cite{stemple:2000}, functions \textit{drop\/} and \textit{raise\/} capture the essence of linguistic reflection. 

For instance, when evaluating \textit{pgm\/} in a state $S$ the result should be a tree value, to which we may apply some tree operators to extract a rule associated with some subtree. As this subtree is a value in $B$ (and thus in $B_\mathit{ext}$) we may apply \textit{raise\/} to it to obtain the ASM rule, which could be executed to determine an update set and to update the state. Analogously, when assigning a new term, e.g. a Boolean term in a branching rule, to a subtree of \textit{pgm\/} the value on the right-hand side must be the result of the function \textit{drop\/}, otherwise the term would be evaluated and a Boolean value would be assigned instead.

The functions \textit{drop\/} and \textit{raise\/} can be applied to function names as well, so they can be used as values stored within \textit{pgm\/} and used as function symbols in rules. In particular, if $\mathcal{O}$ denotes the set of nodes of a tree, then each $o \in \mathcal{O}$ is a value in the base set, but $\textit{raise\/}(o)$ denotes a nullary function symbol that is bound in a state to the subtree $\hat{o}$. However, as it is always clear from the context, when a function name $f$ is used as a value, i.e. as $\textit{drop\/}(f)$, this subtle distinction can be blurred.

Finally, the self-representation as defined above involves several non-logical constants such as the keywords for the rules, which are labels in $L$. For a theoretical analysis it is important to extract from the representation the decisive terms defined over $\Sigma$ and $B$. That is, we further require an {\em extraction function} $\beta: \mathbb{T}_{ext} \rightarrow \bar{\mathbb{T}}$, which assigns to each term included in the extended base set $B_{ext}$ a tuple of terms in $\bar{\mathbb{T}}$ defined over $\Sigma$, $B$ and the background operators. This extraction function $\beta$ on rule terms is easily defined as follows:
\begin{align*}
\beta( \textit{label\_hedge}&(\texttt{update},\texttt{func} \langle f \rangle \texttt{term} \langle t_1 \dots t_n \rangle \texttt{term} \langle t_0 \rangle) ) = \\
& ( \multiset{t_0, t_1, \dots, t_n \mid \text{true}} ) \\
\beta( \textit{label\_hedge}&(\texttt{partial},\texttt{func} \langle f \rangle \texttt{func} \langle op \rangle \texttt{term} \langle t_1 \dots t_n \rangle \texttt{term} \langle t_1^\prime \dots t_m^\prime \rangle) ) = \\
& ( \multiset{ (t_1 ,\dots, t_n, op(f(t_1 ,\dots, t_n), t_1^\prime, \dots, t_m^\prime)) \mid \text{true}} )\\
\beta( \textit{label\_hedge}&(\texttt{par}, \texttt{rule} \langle t_1 \rangle \;\dots\; \texttt{rule} \langle t_k \rangle) ) = \\ 
& (t_1^1 ,\dots, t_1^{n_1}, \dots, t_k^1 ,\dots, t_k^{n_k}) \; \text{for}\; \beta(t_i) =(t_i^1 ,\dots, t_i^{n_i}) \;(1 \le i \le k) \\
\beta( \textit{label\_hedge}&(\texttt{if},\texttt{bool} \langle \varphi \rangle \texttt{rule} \langle t_1 \rangle \texttt{rule} \langle t_2 \rangle) ) = \\
& ( \multiset{\varphi, \mid \text{true}}, \dots, \multiset{ (t^1_{i,1} ,\dots, t^1_{i,n_i}) \mid \varphi_i \wedge \varphi}, \dots, \\
&\qquad\qquad \dots, \multiset{ ( t^2_{j,1} ,\dots, t^2_{j,n_j^\prime}) \mid \psi_j \wedge \neg\varphi}, \dots ) \\
&\text{for}\; \beta(t_1) = ( \dots \multiset{ (t^1_{i,1} ,\dots, t^1_{i,n_i}) \mid \varphi_i }, \dots) \;(1 \le i \le m_1)\\
&\text{and}\; \beta(t_2) = ( \dots \multiset{ (t^2_{j,1} ,\dots, t^2_{j,n_j^\prime}) \mid \psi_j }, \dots ) \; (1 \le j \le m_2)\\
\beta( \textit{label\_hedge}&(\texttt{forall}, \texttt{term}\langle x \rangle, \texttt{bool}\langle\varphi\rangle, \texttt{rule} \langle t \rangle) ) = \\
& ( \dots, \multiset{ (t_{i,1} ,\dots, t_{i,n_i}) \mid \varphi_i \wedge \varphi}, \dots, \multiset{ (t_{i,1} ,\dots, t_{i,n_i}) \mid \varphi_i \wedge \neg\varphi}, \dots )\\
&\text{for}\; \beta(t) = ( \dots \multiset{ (t_{i,1} ,\dots, t_{i,n_i}) \mid \varphi_i }, \dots) \;(1 \le i \le k)\\
\beta( \textit{label\_hedge}&(\texttt{let}, \texttt{term} \langle x \rangle \texttt{term} \langle t \rangle \texttt{rule} \langle t^\prime \rangle) ) = \\
& (t, t_1 ,\dots, t_n) \;\text{for}\; \beta(t^\prime[x \mapsto t]) = (t_1 ,\dots, t_n) \\
\beta( \textit{label\_hedge}&(\texttt{import}, \texttt{term} \langle x \rangle \texttt{rule} \langle t \rangle) ) = \beta(t)
\end{align*}

Then the {\em background of an rASM} is defined by a background class $\mathcal{K}$ over a background signature $V_K$. It must contain an infinite set \textit{reserve\/} of reserve values, the equality predicate, the undefinedness value \textit{undef\/}, and a set of labels
\begin{align*}
L = \{ & \texttt{pgm},  \texttt{signature}, \texttt{rule}, \texttt{func}, \texttt{name}, \texttt{arity}, \texttt{update}, \texttt{if}, \texttt{term}, \texttt{bool}, \\
& \texttt{forall}, \texttt{par}, \texttt{let}, \texttt{partial}, \texttt{import} \} \; .
\end{align*}

The background class must further define truth values and their connectives, tuples and projection operations on them, multisets with union and comprehension operators natural numbers and operations on them, trees in $T_L$ and tree operations, and the function $\mathbf{I}$, where $\mathbf{I} x . \varphi$ denotes the unique $x$ satisfying condition $\varphi$.

The background must further provide functions: $\textit{drop\/}: \hat{\mathbb{T}}_{ext} \rightarrow B_{ext}$ and $\textit{raise\/}: B_{ext} \rightarrow \hat{\mathbb{T}}_{ext}$ for each base set $B$ and extended base set $B_{ext}$, as well as the derived {\em extraction function} $\beta$ defined above. In this definition we use $\hat{\mathbb{T}}_{ext}$ to denote the union of:

\begin{enumerate}

\item The set $\mathbb{T}_{ext}$ of all terms included in $B_{ext}$;

\item The set of all ASM rules which can be formed together with the terms included in $B_{ext}$;

\item The set of all possible signatures of the form signature $\Sigma \cup \Sigma _{\textit{ext\/}}$ such that $\Sigma_{\textit{ext\/}}$ is a finite subset of $\{raise(v_i) \mid v_i \text{ in the reserve}\}$ and each function symbol of  $\Sigma_{\textit{ext\/}}$ has some arbitrary fixed arity. 

\end{enumerate}

\subsection{Reflective ASMs}

\begin{definition}\label{def-rasm}

A {\em reflective ASM} (rASM) $\mathcal{M}$ comprises an (initial) signature $\Sigma$ containing a $0$-ary function symbol \textit{pgm\/}, a background as defined above, and a set $\mathcal{I}$ of initial states over $\Sigma$ closed under isomorphisms such that any two states $I_1, I_2 \in \mathcal{I}$ coincide on \textit{pgm\/}. If $S$ is an initial state, then the signature $\Sigma_S = \textit{raise\/}(\textit{signature\/}(\text{val}_S(\textit{pgm\/})))$ must coincide with $\Sigma$. Furthermore, $\mathcal{M}$ comprises a state transition function $\tau(S)$ on states over the (extended) signature $\Sigma_S$ with $\tau(S) = S + \Delta_{r_S}(S)$, where the rule $r_S$ is defined as $\textit{raise\/}(\textit{rule\/}(\text{val}_S(\textit{pgm\/})))$ over the signature $\Sigma_S = \textit{raise\/}(\textit{signature\/}(\text{val}_S(\textit{pgm\/})))$.

\end{definition}

In Definition \ref{def-rasm} we use extraction functions \textit{rule\/} and \textit{signature\/} defined on the tree representation of a parallel ASM in \textit{pgm\/}. These are simply defined as $\textit{signature\/}(t) = \textit{subtree\/}(\textbf{I} o . \textit{root}(t) \prec_c o \wedge \textit{label\/}(o) = \texttt{signature} )$ and $\textit{rule\/}(t) = \textit{subtree\/}(\textbf{I} o . \textit{root}(t) \prec_c o \wedge \textit{label\/}(o) = \texttt{rule} )$.

Our first main result is that each rASM satisfies the defining postulates, i.e. each rASM defines an RA. This constitutes the plausibility part of our reflective parallel ASM thesis.

\begin{theorem}\label{thm:plausibility}

Every reflective ASM $\mathcal{M}$ defines a RA.

\end{theorem}

\begin{proof}
First consider sequential time. By definition a rASM $\mathcal{M}$ comprises a set $\mathcal{I}$ of initial states defined over an (initial) signature $\Sigma$. Other states are defined through reachability by finitely many applications of the state transition function $\tau$. This gives rise to the set of states $\mathcal{S}$, where each state is defined over a signature $\Sigma_S$ with $\Sigma \subseteq \Sigma_S$. This state transition function $\tau$ is explicitly defined by the rASM. Furthermore, in every initial state $S_0 \in \mathcal{I}$ we have a unique rule $r_{S_0} = \textit{raise\/}(\textit{rule\/}(\text{val}_{S_0}(\textit{pgm\/})))$ using the rule extraction function $\textit{rule\/}$ defined in Subsection \ref{ssec:treerep}.

Concerning the abstract state postulate the required properties (i) and (v) are explicitly built into the definition of an rASM \ref{def-rasm}, and the set $\mathcal{I}$ is closed under isomorphism. This extends to the set $\mathcal{S}$ of all states due to the definition of the state transition function $\tau$. This gives the required properties (ii) and (iii). Concerning property (iv) we have $\Sigma_{alg} = \{ \textit{pgm\/} \}$, so the restriction of a state $S$ is simply given by $\text{val}_S(\textit{pgm\/})$. Applying the functions \textit{rule\/} and \textit{signature\/} from Subsection \ref{ssec:treerep} to this tree value give the rule $r_S$ and the signature $\Sigma_S$, which define the algorithm $\mathcal{A}(S)$ with the desired property.

The requirements for the background postulate are also built into the definition of rASMs. The extraction function $\beta$ has been defined explicitly in Subsection \ref{ssec:treerep}.

Finally, concerning the bounded exploration postulate take $W = \{ \textit{pgm\/} \}$. Let $S$ and $S^\prime$ be two states that strongly coincide on $W$, so we have $\text{val}_S(\textit{pgm\/}) = \text{val}_{S^\prime}(\textit{pgm\/})$. As $r_S = \textit{raise\/}(\textit{rule\/}(\text{val}_S(\textit{pgm\/})))$, we obtain $r_S = r_{S^\prime}$ with signature $\Sigma_S = \textit{raise\/}(\textit{signature\/}(\text{val}_S(\textit{pgm\/}))) = \Sigma_{S^\prime}$. Also, applying the extraction function $\beta$ gives $\beta(\text{val}_S(\textit{pgm\/})) = \beta(\text{val}_{S^\prime}(\textit{pgm\/}))$. Let this tuple of terms be $(t_1 ,\dots, t_n)$. Then the $t_i$ are all the terms over $\Sigma_S$ and the base set $B$ that appear in $r_S$ and $r_{S^\prime}$. In particular, $\{ t_1 ,\dots, t_n \}$ is a bounded exploration witness for the ASM defined by $\Sigma_S$ and $r_S$.

We get $\text{val}_S(\textit{raise\/}(\beta(\textit{pgm\/}))) = \text{val}_{S^\prime}(\textit{raise\/}(\beta(\textit{pgm\/})))$ from the definition of strong coincidence, hence $\text{val}_S(\textit{raise\/}(t_i)) = \text{val}_{S^\prime}(\textit{raise\/}(t_i))$ holds for all $i = 1 ,\dots, n$, i.e. the states $S$ and $S^\prime$ coincide on a bounded exploration witness. Thus, we get $\ddot{\Delta}_{\mathcal{A}}(S) = \ddot{\Delta}_{r_S}(S) = \ddot{\Delta}_{r_{S^\prime}}(S^\prime) = \ddot{\Delta}_{\mathcal{A}}(S^\prime)$, which shows the satisfaction of the bounded exploration postulate and completes the proof.
\end{proof}

\section{The Reflective Parallel ASM Thesis}\label{sec:theory}

Our second main result is the converse of Theorem \ref{thm:plausibility}, i.e. every RA $\mathcal{A}$ can be step-by-step simulated by a behaviourally equivalent rASM $\mathcal{M}$. Hence rASMs capture all RAs regardless how algorithms are represented by terms. This constitutes the characterisation part of the reflective parallel ASM thesis. 

\begin{theorem}\label{thm:characterisation}

For every RA $\mathcal{A}$ there is a behaviourally equivalent rASM $\mathcal{M}$.

\end{theorem}

The proof is done by a lengthy sequence of lemmata. The first part exploits the proof of the parallel ASM thesis \cite{ferrarotti:tcs2016}, and the second part exploits the tree algebra \cite{schewe:scp2022}. Throughout this proof we fix a bounded exploration witness $W$ for a parallel algorithm $A$, and without loss of generality we assume that $W$ is closed under subterms in the following sense: if $\langle t \mid \varphi \rangle \in W$, then also $\langle t ^\prime \mid \exists x_1 ,\dots, x_k \, \varphi \rangle \in W$, where $t^\prime$ is a subterm of $t$ and $\{ x_1 ,\dots, x_k \} = \textit{free}(\varphi) - \textit(t^\prime)$.

Let $W_{st}$ be the subset of $W$ containing ``standard'' terms, i.e. those terms that do not contain function symbols in $\Sigma_{alg}$, and let $W_{pt} = W - W_{st}$ be the complement containing ``program terms''. 

\subsection{Critical Structures}

According to the bounded exploration postulate the update set $\Delta_{\mathcal{A}}(S)$ in a state $S$ is determined by values $\text{val}_S(t)$ resulting from the interpretation of terms $t \in W$ and $t = \pi_i(\beta(\text{val}_S(t^\prime)))$ with $t^\prime \in W_{pt}$ and some projection function $\pi_i$. However, according to the abstract state postulate the successor state $\tau(S)$ is already determined by the algorithm $\mathcal{A}(S)$ represented in the state $S$, i.e. $\Delta_{\mathcal{A}}(S) = \Delta_{\mathcal{A}(S)}(S)$ and therefore, instead of $\text{val}_S(t)$ with $t \in W$ it suffices to consider $t \in W_{st}$---values $\text{val}_S(t)$ with $t \in W_{pt}$ must already be covered by the values $\text{val}_S(\pi_i(\beta(\text{val}_S(t^\prime))))$ with $t^\prime \in W_{pt}$. We use the notation $W_{\beta} = \{ \beta(\text{val}_S(t)) \mid t \in W_{pt} \}$ and $W_S = W_{st} \cup W_{\beta}$. Without loss of generality we can assume that $W_S$ is closed under subterms.

For a state $S$ the terms in $W_S$ are called {\em critical terms} in $S$. The value $\text{val}_S(t)$ of a critical term $t \in W_S$ is called a {\em critical value} in $S$. A {\em critical tuple} is a tuple $(a_0, \dots, a_r)$, where each $a_i$ is a critical value.

In the following we proceed analogously to the proof of the main theorem in \cite{ferrarotti:tcs2016} concerning the capture of parallel algorithms by ASMs, i.e. we start from a single state $S$ and the update set $\Delta_{\mathcal{A}}(S)$ in that state. We can first show that all values appearing in updates in an update set $\Delta_{\mathcal{A}}(S)$ are critical in $S$. The proof of this lemma is analogous to the corresponding proof for parallel ASMs, but it has to be extended to capture also updates that result from aggregation of shared updates.

\begin{lemma}\label{lemma:critical_terms}

If $((f, (v_1, \dots v_n), v_0))$ is an update in $\Delta_{\mathcal{A}}(S)$, then it follows that either $v_0, v_1, \dots, v_n$ are critical values in $S$ or $(v_0, v_1, \dots, v_n)$ is an aggregation tuple built from critical values.

\end{lemma}

\begin{proof}

The update $((f, (v_1, \dots v_n)), v_0)$ may be the result of merging several shared updates or not. In the latter case assume that one value $v_i$ is not critical. Then choose a structure $S_1$ by replacing $v_i$ by a fresh value $b$ without changing anything else. Then $S_1$ is isomorphic to $S$ and thus a state by the abstract state postulate.

Let $t \in W_S$ be a critical term. Then we must have $\text{val}_S(t) = \text{val}_{S_1}(t)$, so $S$ and $S_1$ coincide on $W_S$, so they strongly coincide on $W$. The bounded exploration postulate implies $\Delta_{\mathcal{A}}(S) = \Delta_{\mathcal{A}}(S_1)$ and hence $(f(v_1,\dots,v_n),v_0)$ $\in \Delta_{\mathcal{A}}(S_1)$. However, $v_i$ does not appear in the structure $S_1$ and thus cannot appear in $\Delta_{\mathcal{A}}(S_1)$, which gives a contradiction.

In case the update $((f, (v_1, \dots v_n)), v_0)$ is the result of merging several shared updates we have shared updates $((f, (v_1, \dots v_n)), op_i, v_{i1}^\prime,\dots,v_{im_i}^\prime)$ ($i = 1,\dots,k$), and 
\begin{gather*}
v_0 = op_1(op_2(\dots (op_k(f_S(v_1,\dots,v_n),v_{k1}^\prime, \dots , v_{km_k}^\prime)),\dots, v_{21}^\prime, \\\dots,v_{2m_2}^\prime), v_{11}^\prime,\dots,v_{1m_1}^\prime) \; . 
\end{gather*}
Applying the same argument as in the first case (using $\ddot{\Delta}_{\mathcal{A}}(S) = \ddot{\Delta}_{\mathcal{A}}(S_1)$) we conclude that $(v_0, v_1, \dots, v_n)$ is an aggregation tuple.
\end{proof}

Lemma \ref{lemma:critical_terms} implies that every update in $\Delta_{\mathcal{A}}(S)$ can be produced by an ASM rule, which is either an assignment or the parallel composition of partial updates. For reflective sequential algorithms this suffices to construct a rule $r_S$ producing the whole update set $\Delta_{\mathcal{A}}(S)$. For reflective parallel algorithms, however, it becomes necessary to bundle updates that can be produced by \texttt{forall}-rules. For this we define a purely relational and finite structure $S|_{W}$, which captures the part of the state $S$ that can be accessed with the set $W_S$ of witness terms. 

The \emph{critical (sub)structure} $S|_W$ of $S$ is defined over the vocabulary $\Sigma_{W} = \{R_{\alpha_1}, \ldots, R_{\alpha_m}\}$, where for $1 \leq i \leq m$ and $\alpha_i = \langle (t_{0}, \ldots, t_{n_i}) \mid \varphi_i(x_1, \ldots, x_{r_i}) \rangle \in W_S$, such that the relation symbol $R_{\alpha_i}$ has arity $n_i+2$ and the following interpretation:
\begin{align*}
R^{S|_{W}}_{\alpha_i} &= \{(b_0, \ldots, b_{n_i}, j \cdot b_0 \cdots b_{n_i}  \cdot a_1 \cdots a_{r_i}) \mid  (a_1, \ldots, a_{r_i}) \in B^{r_i}, \\
&\quad {\bf S} \models \varphi_i(x_1, \ldots, x_{r_i})[a_1, \ldots, a_{r_i}] \text{ and }\\
&\quad \text{val}_{S,\mu[x_1 \mapsto a_1, \ldots, x_{r_i} \mapsto a_{r_i}]}(t_0) = b_0, \ldots, \text{val}_{{\bf S},\mu[x_1 \mapsto a_1, \ldots, x_{r_i} \mapsto a_{r_i}]}(t_{n_i}) = b_{n_i}\} \; , 
\end{align*}
where $j \cdot b_0 \cdots b_{n_i}  \cdot a_1 \cdots a_{r_i}$ denotes a value in $B$ such that the set of these values for fixed $b_0, \dots, b_{n_i}$ is in one-one correspondence to the set of tuples $(a_1, \dots, a_{r_i})$ satisfying the required conditions for the evaluation of $\varphi_i$ and $t_k$ ($0 \le k \le n_i$). An element $a_i$ belongs to the domain of $S|_{W}$ iff for some $\alpha_i \in W$ there is a $\bar{a} \in R^{S|_{W}}_{\alpha_i}(\bar{a})$ such that $a_i$ appears in $\bar{a}$.

Critical structures are purely based on the set of witness terms and in particular do not assume the existence of an equality relation. The values $j \cdot b_0 \cdots b_{n_i}  \cdot a_1 \cdots a_{r_i} \in B$ serve the sole purpose of encoding the multiplicities of the elements in the multiset resulting from the evaluation of $\alpha_i$ in $\bf S$, again without including anything outside of what is prescribed by the witness set.  

\subsection{Types}

With critical structures we can restrict ourselves to consider ingthe properties of tuples that are definable in a given logic over \emph{finite relational structures}.  For this, we use the model-theoretic concept of type.  

Let $\cal L$ be a logic, let ${\bf A}$ be a relational structure of vocabulary $\Sigma$, and let $\bar{a} = (a_1,  \ldots,  a_k)$ be  a $k$-tuple over ${\bf A}$. Let ${\cal L}[\Sigma]$ denote the set of formulae in ${\cal L}$ of vocabulary $\Sigma$. Then the $\cal L$-{\em  type} of $\bar{a}$ in $\bf A$,  denoted $tp_{\bf A}^{\cal L}(\bar{a})$, is the set of formulae in ${\cal L}[\Sigma]$ with free variables among $\{x_1, \ldots, x_k\}$ which are satisfied in ${\bf A}$ by any variable assignment assigning the $i$-th component of $\bar{a}$ to the variable $x_i$ ($1 \leq i \leq k$ ), i.e.
\[ tp_{\bf A}^{\cal L}(\bar{a}) = \{\varphi \in {\cal L}[\Sigma] \mid \mathit{free}(\varphi) \subseteq
\{x_1, \ldots, x_k\} \textrm{ and } {\bf A} \models \varphi[a_1, \ldots, a_k] \}. \]

Note that the ${\cal L}$-type of a given tuple $\bar{a}$ over a relational structure ${\bf A}$ includes not only the properties of all sub-tuples of $\bar{a}$, but also the set of all sentences in ${\cal L}$ which are true when evaluated on ${\bf A}$.

In particular,  we are interested in the properties of tuples which are definable in first-order logic with and without equality (denoted $\mathrm{FO}$ and $\mathrm{FO}_{wo=}$, respectively), i.e., we are interested in $\mathrm{FO}$-types and $\mathrm{FO}_{wo=}$-types, respectively.  $\mathrm{FO}$-types are also known as isomorphism types, because every tuple can be characterized up to isomorphism by its $\mathrm{FO}$-type. $\mathrm{FO}_{wo=}$-types correspond to a (weaker) type of equivalence relation among tuples \cite{casanovas:ndjfl1996}. Instead of the partial isomorphism condition used in the Ehrenfeucht-Fra\"{\i}ss\'e characterization of $\mathrm{FO}$ equivalence, the Ehrenfeucht-Fra\"{\i}ss\'e characterization of $\mathrm{FO}_{wo=}$ equivalence involves the following condition:

Let $\bf A$ and $\bf B$ be relational structures of some vocabulary $\Sigma$. A relation $p \subseteq A \times B$ is a \emph{partial relativeness correspondence} iff for every $n$-ary relation symbol $R \in \Sigma$ and every $(a_1, b_1), \ldots, (a_n, b_n) \in p$ we have $(a_1, \ldots, a_n) \in R^{\bf A}$ iff $(b_1, \ldots, b_n) \in R^{\bf B}$.
Then $\bf A$ and $\bf B$ are \emph{$m$-finitely relative} via $(I_k)_{k \leq m}$ (denoted $(I_k)_{k \leq m}: {\bf A} \sim_m {\bf B}$) iff the following holds:

\begin{enumerate}

\item Every $I_k$ is a nonempty set of partial relativeness correspondences.

\item For any $k + 1 \leq m$, any $p \in I_{k+1}$ and any $a \in A$, there are $b \in B$ and $q \in I_k$ such that $q \supseteq p$ and $(a,b) \in q$ (forth condition).

\item For any $k + 1 \leq m$, any $p \in I_{k+1}$ and any $b \in B$, there are $a \in A$ and $q \in I_k$ such that $q \supseteq p$ and $(a,b) \in q$ (back condition).

\end{enumerate}

Then the following theorem is an immediate consequence of \cite[Prop.~4.5, Thm.~4.6]{casanovas:ndjfl1996}.

\begin{theorem}\label{FOwo=Characterization}

Let ${\bf A}$ and ${\bf B}$ be relational structures of some vocabulary $\Sigma$. For every $r \geq 0$, $r$-tuples $\bar{a} = (a_1, \ldots, a_r) \in A^r$ and $\bar{b}=(b_1, \ldots, b_r) \in B^r$, and $p = \{(a_1, b_1), \ldots, (a_r, b_r)\}$, the following holds:

\begin{enumerate}

\item There is a sequence $(I_k)_{k \leq m}$ such that $(I_k)_{k \leq m}:{\bf A} \sim_m {\bf B}$ and $p \in I_m$ iff for every equality-free formula $\varphi$ of quantifier rank up to $m$ with at most $r$ distinct free variables ${\bf B} \models \varphi[\bar{b}]$ holds iff ${\bf A} \models \varphi[\bar{a}]$ holds.

\item For every $m \geq 0$ there is a sequence $(I_k)_{k \leq m}$ such that $(I_k)_{k \leq m}:{\bf A} \sim_m {\bf B}$ and $p \in I_m$ iff for every equality-free formula $\varphi$ with at most $r$ distinct free variables ${\bf B} \models \varphi[\bar{b}]$ holds iff ${\bf A} \models \varphi[\bar{a}]$ holds.

\end{enumerate}

\end{theorem}

\subsection{Indistinguishable Updates}

We want to show that if $\bar{a} = (a_0,\dots,a_r)$ is a critical tuple that defines an update $((f,(a_1,\dots,a_r)),a_0)$ in some update set of the reflective parallel algorithm $A$ and $\bar{b}$ has the same ${\cal L}$-type over the critical structure of $A$, then also $\bar{b}$ is a critical tuple that defines an update $((f,(b_1,\dots,b_r)),b_0)$ in the same update set. This will lead to our Lemma~\ref{also_in_update_set}. For the proof it will turn out to be convenient, if there exists an isomorphism that takes $\bar{a}$ to $\bar{b}$. However, this cannot always be guaranteed. Therefore, in this subsection we will show how the general case can be reduced to the specific case assuming such an isomorphism, and the latter case will be handled in the next subsection.

Thus, for every parallel algorithm $A$ we define next a modified parallel algorithm $A^*$ by means of a bijection from the states of $A$ to states of $A^*$. Then we prove that this modified version $A^*$ of $A$ satisfies the properties that are required for the proof of Lemma~\ref{also_in_update_set}.

For each state $S_i$ of a reflective parallel algorithm $A$, let $S_i^*$ denote a corresponding state of vocabulary $\Sigma^* = \{g_i \mid f_i \in \Sigma\} \cup \{h\}$ such that:

\begin{itemize}

\item The base set $B_i^*$ of $S_i^*$ is the disjoint union of the base set $B_i$ of $S_i$ with the positive natural numbers $\mathbb{N}^+$.

\item $h^{S_i^*}$ is a static function that is a bijection from $B_i$ to $\mathbb{N}^+$.

\item For every $r$-ary function $g_j \in \Sigma^*$ and every $(n+1)$-ary tuple $(c_0, c_1, \ldots, c_n)$ in $(B_i^*)^{n+1}$ we have
\[ g_j^{S^*}(c_1, \ldots, c_n) = 
\begin{cases}
  c_0 & \text{if there is a} \; (d_0, d_1, \ldots, d_n) \in (B_i)^{n+1}  \\ & \quad \text{such that} f^{S}_j(d_1, \ldots, d_n) = d_0 \; \text{and} \\ & \quad c_i = \mathit{prime}(i+1)^{h(d_i)} \; \text{for} \; 0 \leq i \leq n; \\
\texttt{false} & \text{if the previous condition does not hold and}\;\\ &\quad f \; \text{is marked as relational};\\
\texttt{undef} & \text{otherwise.}
\end{cases}
\]
Here, $\mathit{prime}(i)$ denotes the $i$-th prime number in the sequence of primes.  

\end{itemize}

We define $A^*$ as the parallel algorithm with set of states ${\cal S}_{A^*} = \{ S_i^* \mid S_i \in {\cal S}_A\}$, set of initial states ${\cal I}_{A^*} = \{ S_i^* \mid S_i \in {\cal I}_A\}$ and transition function $\tau_{A^*}$ such that for every $S^*_i, S^*_j \in {\cal S}_{A^*}$ it holds that:

\begin{enumerate}
\item If the base sets of $S^*_i$ and $S^*_j$ coincide, then $h^{S_i^*} = h^{S_j^*}$.

\item $\tau_{A*}(S^*_i) = S^*_j$ iff $\tau_{A}(S_i) = S_j$. 

\end{enumerate}

Let $t$ be a term of vocabulary $\Sigma$. We define $t^*$ as the term of vocabulary $\Sigma^*$ obtained from $t$ as follows:

\begin{itemize}

\item If $t$ is a nullary function symbol $f_i \in \Sigma$, then $t^*$ is $g_i$. 

\item If $t$ is a variable $x_i \in V$, then $t^*$ is $2^{h(x_i)}$.

\item If $t$ is of the form $f_i(t_1, \ldots, t_r)$ where $f_i \in \Sigma$, $\mathit{arity}(f_i) = r$ and $t_1, \ldots, t_r$ are terms, then $t^*$ is $g_i(\mathit{prime}(2)^{\log_2(t^*_1)}, \ldots, \mathit{prime}(r+1)^{\log_2(t^*_r)})$.

\item If $t(\bar{y})$ is a multiset comprehension term of the form $\langle s(\bar{x}, \bar{y}) \mid \varphi(\bar{x}, \bar{y}) \rangle_{\bar{y}}$, then $t^*$ is $2^{h(\langle h^{-1}(\log_2(s^*(\bar{x}, \bar{y}))) \mid h^{-1}(\log_2(\varphi^*(\bar{x}, \bar{y})))\rangle_{\bar{y}})}$.

\end{itemize}

\begin{lemma}\label{similarresult}

Let $S$ be a state of vocabulary $\Sigma$, $\alpha$ a term of vocabulary $\Sigma$, $\alpha^*$ the corresponding term of vocabulary $\Sigma^*$ as defined above, and let $\mu$ be a variable assignment over $S$. We have that $\mathrm{val}_{S, \mu}(\alpha) = a$ iff $\mathrm{val}_{S^*, \mu}(\alpha^*) = 2^{h(a)}$.

\end{lemma}

\begin{proof}

We proceed by induction on $\alpha$. 

If $\alpha$ is a nullary function symbol $f_i \in \Sigma$, then $\mathrm{val}_{{\bf S}, \mu}(f_i) = f^{\bf S}_i$ and $\mathrm{val}_{{\bf S}^*, \mu}(\alpha^*) = g^{\bf S^*}_i$, and by definition $f^{\bf S}_i  = a$ iff $g^{\bf S^*}_i = 2^{h(a)}$.

If $\alpha$ is a variable $x_i$, then $\mathrm{val}_{{\bf S},\mu}(x_i) = \mu(x_i)$ and $\mathrm{val}_{{\bf S}^*,\mu}(\alpha^*) = 2^{h(\mu(x_i))}$, and clearly, $\mu(x_i) = a$ iff $2^{h(\mu(x_i))} = 2^{h(a)}$.

If $\alpha$ is of the form $f_i(t_1, \ldots, t_r)$ where $f_i$ is an $r$-ary function symbol in $\Sigma$ and $t_1, \ldots, t_r$ are terms of vocabulary $\Sigma$, then by induction hypothesis $\mathrm{val}_{{\bf S}, \mu}(t_i) = a_i$ iff $\mathrm{val}_{{\bf S}^*, \mu}(t_i^*) = 2^{h(a_i)}$ for every $1 \leq i \leq r$. Thus, by definition we get that $\mathrm{val}_{{\bf S}, \mu}(\alpha) = f_i^{\bf S}(a_1, \ldots, a_r) = a$ iff 
\[ \mathrm{val}_{{\bf S}^*, \mu}(\alpha^*) = g_i^{{\bf S}^*}(\mathit{prime}(2)^{\log_2(2^{h(a_1)})}, \ldots, \mathit{prime}(r+1)^{\log_2(2^{h(a_r)})}) = 2^{h(a)} \; . \]

If $\alpha$ is of the form $\langle s(\bar{x}, \bar{y}) \mid \varphi(\bar{x}, \bar{y}) \rangle_{\bar{y}}$ where $\bar{x} = (x_1, \ldots, x_n)$ and $\bar{y} = (y_1, \ldots, y_m)$, then for every $\bar{b} \in S^n$ it holds by induction hypothesis that $\mathrm{val}_{{\bf S}, \mu[\bar{x} \mapsto \bar{b}]}(s(\bar{x}, \bar{y})) = a_i$ and $\mathrm{val}_{{\bf S}, \mu[\bar{x} \mapsto \bar{b}]}(\varphi(\bar{x}, \bar{y})) = a_j$ iff $\mathrm{val}_{{\bf S}^*, \mu[\bar{x} \mapsto \bar{b}]}(s^*(\bar{x}, \bar{y})) = 2^{h(a_i)}$ and $\mathrm{val}_{{\bf S}^*, \mu[\bar{x} \mapsto \bar{b}]}(\varphi^*(\bar{x}, \bar{y})) = 2^{h(a_j)}$. Thus $\mathrm{val}_{{\bf S},\mu}(\alpha) = a$ iff
\[ \mathrm{val}_{{\bf S}^*,\mu}(\langle h^{-1}(\log_2(s^*(\bar{x}, \bar{y}))) \mid h^{-1}(\log_2(\varphi^*(\bar{x}, \bar{y}))) \rangle_{\bar{y}}) = a \]
iff $\mathrm{val}_{{\bf S}^*,\mu}(\alpha^*) = 2^{h(a)}$.
\end{proof}

\begin{lemma}\label{lemma:properties}

Let $A$ be a reflective parallel algorithm and $W$ be a bounded exploration witness for $A$. The following holds:

\begin{enumerate}

\item For every $S_i \in {\cal S}_{A}$ and every $f_i \in \Sigma$ we have $(f_i, (d_1, \ldots, d_n), d_0) \in \Delta_A(S_i)$ iff
\[(g_i, (\mathit{prime}(2)^{h(d_1)}, \ldots, \mathit{prime}(n+1)^{h(d_n)}), \mathit{prime}(1)^{h(d_0)}) \in \Delta_{A^*}(S^*_i) \; .\]

\item The set $W^* = \{\alpha^*_i \mid \alpha_i \in W\}$ of witness terms is a bounded exploration witness for the modified algorithm $A^*$.\label{witnessSet*}

\end{enumerate}

\end{lemma} 

\begin{proof}

Let $\tau_A(S_i) = S_j$. As $(f_i, (d_1, \ldots, d_n), d_0) \in \Delta_A(S_i)$, we can assume $f_i^{S_j}(d_1, \ldots, d_b) = d_0$ and $f_i^{S_i}(d_1, \ldots, d_b) \neq d_0$. Hence 
\[ g_i^{S^*_j}(\mathit{prime}(2)^{h(d_1)}, \dots, \mathit{prime}(n+1)^{h(d_n)}) = \mathit{prime}(1)^{h(d_0)} \]
and $g_i^{S^*_i}(\mathit{prime}(2)^{h(d_1)}, \ldots,\mathit{prime}(n+1)^{h(d_n)}) \neq \mathit{prime}(1)^{h(d_0)}$ hold. It follows that
\[ (g_i, (\mathit{prime}(2)^{h(d_1)}, \ldots, \mathit{prime}(n+1)^{h(d_n)}), \mathit{prime}(1)^{h(d_0)}) \in \Delta_{A^*}(S^*_i) \; . \]
The same argument can be used to prove the other direction of condition (i).

For condition (ii) let us assume that $S^*_1$ and $S^*_2$ are states of $A^*$ that coincide on $W^*$ and $\Delta_{A^*}(S^*_1) \neq \Delta_{A^*}(S^*_2)$ holds. As $S^*_1$ and $S^*_2$ coincide on $W^*$, it follows from Lemma~\ref{similarresult} and the construction of $W^*$ from $W$ that $S_1$ and $S_2$ coincide on $W$. Given that $W$ is a bounded exploration witness for $A$, it follows from the bounded exploration postulate that  $\Delta_{A}(S_1) = \Delta_{A}(S_2)$. Hence also  $\Delta_{A^*}(S^*_1) = \Delta_{A^*}(S^*_2)$ holds, which contradicts our assumption.
\end{proof}

\subsection{A Key Lemma}

The following key lemma shows that updates composed by tuples that share a same $\mathrm{FO}_{wo=}$-type in a critical structure are indistinguishable by the algorithm from one another. This implies that if two tuples that define two different updates with the same dynamic function share the same $\mathrm{FO}_{wo=}$-type, then either both updates occur together in an update set or neither of them does.

\begin{lemma}\label{also_in_update_set}

Let $A$ be a reflective parallel algorithm, let $\bf S$ be a state, $\bar{a} = (a_0, \ldots, a_r)$, and let $(f,(a_1, \ldots, a_r), a_0) \in \Delta_A(S)$. For every $(r+1)$-tuple of critical values $\bar{b} = (b_0, \ldots, b_r)$ with $\mathit{tp}^{\mathrm{FO}_{wo=}}_{S|_{W}}(\bar{b}) = \mathit{tp}^{\mathrm{FO}_{wo=}}_{S|_{W}}(\bar{a})$ the update $(f, (b_1, \ldots, b_r), b_0)$ also belongs to $\Delta_A(S)$.

\end{lemma}

\begin{proof}

By contradiction. Assume that $(f, (b_1, \ldots, b_r), b_0) \not\in \Delta_A(S)$. Let $S^*$ be the state of the modified parallel algorithm $A^*$ corresponding to $S$, and let $J^*$ be the state isomorphic to $S^*$ induced by the automorphism $\zeta$ of $S^*$ such that $\zeta(x) = \mathit{prime}(i+1)^{h(b_i)}$ if $x$ is $\mathit{prime}(i+1)^{h(a_i)}$ for some $0 \leq i \leq r$, $\zeta(x) = \mathit{prime}(i+1)^{h(a_i)}$ if $x$ is $\mathit{prime}(i+1)^{h(b_i)}$ for some $0 \leq i \leq r$, and $\zeta(x) = x$ otherwise.  

By the Abstract State postulate, $J^*$ is also a state of $A^*$. By part (i) of Lemma~\ref{lemma:properties} we have $(g, (\mathit{prime}(2)^{h(a_1)}, \ldots, \mathit{prime}(r+1)^{h(a_r)}), \mathit{prime}(1)^{h(a_0)}) \in \Delta_{A^*}(S^*)$. Hence by means of the isomorphism $\zeta$ we get
\[ (g, (\mathit{prime}(2)^{h(b_1)}, \ldots, \mathit{prime}(r+1)^{h(b_r)}), \mathit{prime}(1)^{h(b_0)}) \in \Delta_{A^*}(J^*) \; . \]

We claim that $S^*$ and $J^*$ strongly coincide on $W^*$. Then, by part~(ii) of Lemma~\ref{lemma:properties} and the bounded exploration postulate, we get that $\Delta_{A^*}(S^*) = \Delta_{A^*}(J^*)$, but also
$(f, (\mathit{prime}(2)^{h(b_1)}, \ldots, \mathit{prime}(r+1)^{h(b_r)}), \mathit{prime}(1)^{h(b_0)}) \in \Delta_{A^*}(S^*)$. Consequently, $(f, (b_1, \ldots, b_r), b_0) \in \Delta_A(S)$ and part~(i) of Lemma~\ref{lemma:properties} gives a contradiction.

To finalize the proof we need to show our claim. From the characterization in Theorem~\ref{FOwo=Characterization} and the fact that $\mathit{tp}^{\mathrm{FO}_{wo=}}_{S|_{W}}(\bar{b}) = \mathit{tp}^{\mathrm{FO}_{wo=}}_{S|_{W}}(\bar{a})$, we get that for every $m \geq 0$, there is a sequence of partial relativeness correspondences $(I_k)_{k \leq m} : S|_W \sim_m {\bf S}|_W$ with $\{(a_0, b_0), \ldots, (a_r, b_r)\} \in I_m$. Thus, for every $k \leq m$,  every $p_i \in I_k$, every $n$-ary relation symbol $R$ in the relational vocabulary of $S|_W$ and every $(c_1,d_1), \ldots, (c_n, d_n) \in p_i$, it holds that $(c_1, \ldots, c_n) \in R^{S|_W}$ iff $(d_1, \ldots, d_n) \in R^{S|_W}$. By construction of $S|_{W^*}$ and part (i) of Lemma~\ref{lemma:properties}  there is an $x$ such that for every $y$ we have
\begin{gather*}
(\mathit{prime}(1)^{h(c_1)}, \ldots, \mathit{prime}(n-1)^{h(c_{n-1})}, x) \in R^{S^*|_{W^*}} \; \text{iff} \\(\mathit{prime}(1)^{h(d_1)}, \ldots, \mathit{prime}(n-1)^{h(d_{n-1})}, y) \in R^{S^*|_{W^*}} \; .
\end{gather*}

For the same reason there is a $y$ such that for every $x$ the equivalence above holds. By construction of $S^*|_{W^*}$, for every $c'$ and $d'$ such that 
\begin{gather*}
(\mathit{prime}(1)^{h(c_1)}, \ldots, \mathit{prime}(n-1)^{h(c_{n-1})}, c') \in R^{S^*|_{W^*}} \;\text{and}\\
(\mathit{prime}(1)^{h(d_1)}, \ldots, \mathit{prime}(n-1)^{h(d_{n-1})}, d') \in R^{S^*|_{W^*}}
\end{gather*}
hold we have that 
\[ \{(\mathit{prime}(1)^{h(c_1)}, \mathit{prime}(1)^{h(d_1)}) \ldots, (\mathit{prime}(n)^{h(c_n)},\mathit{prime}(n)^{h(d_n)}), (c',d')\} \]
is a partial function, which defines a partial automorphism on $S^*|_{W^*}$. This implies that for every $(I_k)_{k \leq m} : S|_W \sim_m {\bf S}|_W$ with $\{(a_0, b_0), \ldots, (a_r, b_r)\} \in I_m$ and $m \geq 0$ we can build a sequence $I^*_0, \ldots, I^*_m$ of partial automorphisms on $S^*|_{W^*}$ with the back and forth properties and such that 
\[ \{(\mathit{prime}(1)^{h(a_0)}, \mathit{prime}(1)^{h(b_0)}) \ldots, (\mathit{prime}(r+1)^{h(a_r)},\mathit{prime}(r+1)^{h(b_r)})\} \in I^*_m \; . \]

Then by the classical characterization of first-order logic in terms of sequences of partial isomorphisms, we get that $\mathit{tp}^{\mathrm{FO}}_{S^*|_{W^*}}(\bar{b}^*) = \mathit{tp}^{\mathrm{FO}}_{S^*|_{W^*}}(\bar{a}^*)$ for $\bar{b}^* = (\mathit{prime}(1)^{h(b_0)},$ $\ldots, \mathit{prime}(r+1)^{h(b_r)})$ and $\bar{a}^* = (\mathit{prime}(1)^{h(a_0)}, \ldots,$ $\mathit{prime}(r+1)^{h(a_r)})$. 

Now, we proceed by contradiction. Let us assume that there is an 
\[ \alpha_i = \langle (t_0, \ldots, t_{n}) \mid \varphi(x_1, \ldots, x_m) \rangle \in W^* \; \text{such that } \; \text{val}_{{\bf S}_{{\bf I}^*}}(\alpha_i) \neq \text{val}_{{\bf S}_{{\bf J}^*}}(\alpha_i).\] 
Then there is a tuple $\bar{c} = (c_0, \ldots, c_n) \in (S^*)^{n+1}$ (hence also in $(J^*)^{n+1}$) such that either $\text{Mult}(\bar{c}, \text{val}_{S^*}(\alpha_i)) > \text{Mult}(\bar{c}, \text{val}_{J^*}(\alpha_i))$ or $\text{Mult}(\bar{c}, \text{val}_{S^*}(\alpha_i)) < \text{Mult}(\bar{c}, \text{val}_{J^*}(\alpha_i))$.

Let us assume that $\text{Mult}(\bar{c}, \text{val}_{S^*}(\alpha_i)) > \text{Mult}(\bar{c}, \text{val}_{J^*}(\alpha_i))$ and define
\begin{gather*}
A = \{(d_1, \ldots, d_{m}) \in (S^*)^{m} \mid S^* \models \varphi(x_1, \ldots, x_{m})[d_1, \ldots, d_{m}] \; \text{and}\\ \text{val}_{S^*,\mu[x_1 \mapsto d_1, \ldots, x_{m} \mapsto d_{m}]}(t_0) = c_0, \dots, \text{val}_{S^*,\mu[x_1 \mapsto d_1, \ldots, x_{m} \mapsto d_{m}]}(t_n) = c_n \} \; .
\end{gather*}
Analogously, define
\begin{gather*}
B = \{(d_1, \ldots, d_{m}) \in (J^*)^{m} \mid J^* \models \varphi(x_1, \ldots, x_{m})[d_1, \ldots, d_{m}] \;\text{and}\\
\text{val}_{J^*,\mu[x_1 \mapsto d_1, \ldots, x_{m} \mapsto d_{m}]}(t_0) = c_0, \dots, \text{val}_{J^*,\mu[x_1 \mapsto d_1, \ldots, x_{m} \mapsto d_{m}]}(t_n) = c_n \} \; .
\end{gather*}

As $|A| > |B|$ and $S^* \simeq J^*$, there must be some tuple $(d_1, \ldots, d_{m}) \in A$ such that $(\zeta(d_1), \ldots, \zeta(d_{m})) \not\in B$. Furthermore, as $S^* \models \varphi(x_1, \ldots, x_{m})[d_1, \ldots, d_{m}]$ iff $J^*\models \varphi(x_1, \ldots, x_{m})[\zeta(d_1), \ldots, \zeta(d_{m})]$, it follows that 
\begin{gather*}
(\text{val}_{S^*,\mu[x_1 \mapsto d_1, \ldots, x_{m} \mapsto d_{m}]}(t_0), \ldots, \text{val}_{S^*,\mu[x_1 \mapsto d_1, \ldots, x_{m} \mapsto d_{m}]}(t_n)) = \bar{c} \neq \\
\zeta(\bar{c}) = (\text{val}_{J^*,\mu[x_1 \mapsto \zeta(d_1), \ldots, x_{m} \mapsto \zeta(d_{m})]}(t_0), \ldots, \text{val}_{J^*,\mu[x_1 \mapsto \zeta(d_1), \ldots, x_{m} \mapsto \zeta(d_{m})]}(t_n))
\end{gather*}
holds.

Then, as $\zeta$ is the identity function on the set of elements that do not appear in $\bar{a}^*$ or $\bar{b}^*$, we know that there is at least one $c_i$ that appears in $\bar{a}^*$ or $\bar{b}^*$ and such that $\zeta(c_i) \neq c_i$. Without loss of generality let this be the case for exactly one $c_i$ and that $c_i = c_0 = a_0^* = \mathit{prime}(1)^{h(a_0)}$. Then let
\begin{gather*}
\psi(y_0, \ldots y_n) \equiv \exists \bar{z}_1 \ldots \bar{z}_{|A|} \bigg( \bigwedge_{1 \leq j < k \leq |A|} \bar{z}_j \neq \bar{z}_k \wedge \\
\bigwedge_{1 \leq j \leq |A|} (\varphi[\bar{z}_j] \wedge t_0[\bar{z}_j] = y_0 \wedge \cdots \wedge t_n[\bar{z}_j] = y_n) \wedge \\
\neg \exists \bar{z}' \Big( \bigwedge_{1 \leq j \leq |A|} \bar{z}' \neq \bar{z}_j \wedge \varphi[\bar{z}'] \wedge t_0[\bar{z}'] = y_0 \wedge \cdots \wedge t_n[\bar{z}'] = y_n \Big) \bigg) \; ,
\end{gather*}
where for $z_j = (z_{j1}, \ldots, z_{jm})$, we use $\varphi[\bar{z}_j]$ and $t_0[\bar{z}_j], \ldots, t_n[\bar{z}_j]$ to denote the formula and the terms obtained by replacing in $\varphi$ and $t_0, \ldots, t_n$, respectively, every occurrence of a variable $x_i \in \{x_1, \ldots, x_m\}$ by $z_{ji}$. Likewise, we use $\bar{z}_j \neq \bar{z}_k$ to denote the formula $\neg(z_{j1} = z_{k1} \wedge \cdots \wedge z_{jm} = z_{km})$.
 
It follows that $S \models \psi(y_0, \ldots, y_n)[a_0^*, c_1, \ldots, c_n]$ and $J^* \not\models \psi(y_0, \ldots, y_n)$ $[a_0^*, c_1, \ldots, c_n]$. Furthermore, as $\zeta^{-1}(a_0^*) = b_0^* = \mathit{prime}(1)^{h(b_0)}$ holds, we get that $S^* \not\models \psi(y_0, \ldots, y_n)[b_0^*, \zeta^{-1}(c_1), \ldots, \zeta^{-1}(c_n)]$.
But then, for
\begin{gather*}
\psi'(y_0, \ldots y_n) \equiv \exists z_1 \ldots z_{|A|} \bigg( \bigwedge_{1 \leq j < k \leq |A|} z_j \neq z_k \wedge \bigwedge_{1 \leq j \leq |A|} R_{\alpha_i}(y_0, \ldots, y_n, z_j) \wedge \\
\neg \exists z' \Big( \bigwedge_{1 \leq j \leq |A|} z' \neq z_j \wedge R_{\alpha_i}(y_0, \ldots, y_n, z')\Big) \bigg)
\end{gather*}
we get $S^*|_{W^*} \models \psi'(y_0, \ldots, y_n)[a_0^*, c_1, \ldots, c_n]$ and 
\[ S^*|_{W^*} \not\models \psi'(y_0, \ldots, y_n)[b_0^*, \zeta^{-1}(c_1), \ldots, \zeta^{-1}(c_n)] \; , \]
which contradicts the fact that $\mathit{tp}^{\mathrm{FO}}_{S^*|_{W^*}}(\bar{b}^*) = \mathit{tp}^{\mathrm{FO}}_{S^*|_{W^*}}(\bar{a}^*)$. 

By symmetry, if we assume that $\text{Mult}(\bar{c}, \text{val}_{S^*}(\alpha_i)) < \text{Mult}(\bar{c}, \text{val}_{J^*}(\alpha_i))$, the same contradiction is obtained. Hence we must have $\text{Mult}(\bar{c}, \text{val}_{S^*}(\alpha_i)) = \text{Mult}(\bar{c}, \text{val}_{J^*}(\alpha_i))$ which contradicts our assumption that there is an $\alpha_i \in W$ such that $\text{val}_{S^*}(\alpha_i) \neq \text{val}_{J^*}(\alpha_i)$.
\end{proof}

\subsection{Isolating Formulae}

Although types are infinite sets of formulae, a single $\mathrm{FO}_{wo=}$-formula is equivalent to the $\mathrm{FO}_{wo=}$-type of a tuple over a given finite relational structure. The equivalence holds for all finite relational structures of the same schema.

\begin{lemma}\label{isolating_formula}

For  every relational vocabulary $\Sigma$ with no constants, every finite structure ${\bf A}$ of schema $\Sigma$, every $r \geq 0$ and every $r$-tuple $\bar{a}$ over ${\bf A}$ there is a formula $\chi \in \mathit{tp}_{\bf A}^{\mathrm{FO}_{wo=}}(\bar{a})$  such that for any finite relational structure ${\bf B}$ of schema $\Sigma$ and every $r$-tuple $\bar{b}$ over ${\bf B}$ we have ${\bf B}  \models \chi[\bar{b}]$ iff $\mathit{tp}_{\bf A}^{\mathrm{FO}_{wo=}}(\bar{a}) = \mathit{tp}_{\bf B}^{\mathrm{FO}_{wo=}}(\bar{b})$.

\end{lemma}

\begin{proof}  
We define for every $m \in \mathbb{N}$, a formula $\varphi^m_{\bar{a}}$ with free variables $\bar{x} = (x_1, \ldots, x_r)$ and such that ${\bf A} \models \varphi^m_s[\bar{a}]$, which characterizes $\bar{a}$ completely up to equivalence on $\mathrm{FO}_{wo=}$ formulae with quantifier rank $\leq m$. The $\varphi^m_{\bar{a}}$ are defined inductively as follows:    
\begin{align}
\varphi^0_{\bar{a}}(\bar{x}) \equiv& \bigwedge \{\varphi(\bar{x}) \mid \varphi \; \text{is an equality free atomic or negated atomic formula} \notag\\  
& \quad \quad \text{such that} \; {\bf A} \models \varphi[\bar{a}] \}  \label{varphiE} \\
\varphi^{m+1}_{\bar{a}}(\bar{x}) \equiv& \; \bigwedge_{a \in A} \exists x_{r+1} (\varphi_{\bar{a}a}^{m}(\bar{x}, x_{r+1})) \; \wedge \; \forall x_{r+1} \Big( \bigvee_{a \in A}\varphi_{\bar{a}a}^{m}(\bar{x}, x_{r+1}) \Big). \label{varphiA}
\end{align}
We prove first that ${\bf B} \models \varphi^m_{\bar{a}}[\bar{b}]$ iff there is a sequence $(I_k)_{k \leq m}$ such that
\begin{gather}
(I_k)_{k \leq m}:{\bf A} \sim_m {\bf B} \; \text{and} \; p = \{(a_1,b_1), \ldots, (a_r, b_r)\} \in I_m \; . \label{equivalence}
\end{gather}

By part~(i) of Theorem~\ref{FOwo=Characterization} the existence of $(I_k)_{k \leq m}:{\bf A} \sim_m {\bf B}$ with $p \in I_m$ implies that for every equality-free formula $\varphi$ of quantifier rank $\leq m$,  ${\bf B} \models \varphi[b_1, \ldots, b_r]$ iff ${\bf A} \models \varphi[a_1, \ldots, a_r]$. As the quantifier rank of $\varphi^m_{\bar{a}}$ is $m$ and ${\bf A} \models \varphi^m_{\bar{a}}[\bar{a}]$ holds by construction) we get ${\bf B} \models \varphi^m_{\bar{a}}[\bar{b}]$. 

The converse is proven by induction on $m$. For $m=0$ we have ${\bf B} \models \varphi^0_{\bar{a}}[b_1, \ldots, b_r]$, hence $p = \{(a_1,b_1), \ldots, (a_r, b_r)\}$ is a partial relativeness correspondence, and $I_0 = \{p\}$ is a non-empty set of partial relativeness correspondences, even if $\bar{a} = ()$, because $\Sigma$ has no constants and thus $p$ is an empty relation.

For arbitrary $M$ such that ${\bf B} \models \varphi^{m+1}_{\bar{a}}[\bar{b}]$ holds, we have:

\begin{itemize}

\item For every $a \in A$, there is a $b \in B$ such that  ${\bf B} \models \varphi^{m}_{\bar{a}a}[\bar{b}b]$ (by part~(\ref{varphiE}) in the definition of $\varphi^{m+1}_{\bar{a}}$). 

\item For every $b \in B$, there is an $a \in A$ such that  ${\bf B} \models \varphi^{m}_{\bar{a}a}[\bar{b}b]$ (by part~(\ref{varphiA}) in the definition of $\varphi^{m+1}_{\bar{a}}$). 

\end{itemize}

Let $I_{m+1} = \{ p \}$. By induction we get:

\begin{itemize}

\item For every $a \in A$, there is a $b \in B$ and a sequence $(I^{\bar{a}a}_k)_{k\leq m}$ such that $(I^{\bar{a}a}_k)_{k\leq m} : {\bf A} \sim_m {\bf B}$ and  $p \cup \{(a,b)\} \in I^{\bar{a}a}_{m}$.

\item For every $b \in B$, there is an $a \in A$ and a sequence $(I^{\bar{b}b}_k)_{k\leq m}$ such that $(I^{\bar{b}b}_k)_{k\leq m} : {\bf A} \sim_m {\bf B}$ and  $p \cup \{(a,b)\} \in I^{\bar{b}b}_{m}$.

\end{itemize}

Let $I_j = \bigcup_{a \in A} I^{\bar{a}a}_j \; \cup \; \bigcup_{b \in B} I^{\bar{b}b}_j$ (for $0 \leq j \leq m$). In general, $m$-finite relativeness is preserved under this type of element-wise union. Thus we only need to check that the back and forth conditions hold for $I_{m+1}$ and $I_{m}$, i.e. we need to check:

\begin{itemize}

\item For every $a \in A$ there are $b \in B$ and $q \in I_{m}$ such that $q \supseteq p$ and $(a,b) \in q$ (forth condition).

\item For every $b \in B$ there are $a \in A$ and $q \in I_{m}$ such that $q \supseteq p$ and $(a,b) \in q$ (back condition).

\end{itemize}

These properties follow from parts~(\ref{varphiE}) and~(\ref{varphiA}) in the definition of $\varphi^{m+1}_{\bar{a}}$, respectively.

Let $\bar{c} \in A^n$ for some $n \geq 0$. Let $X^m_{\bar{c}} = \{ \bar{d} \in A^n \mid {\bf A} \models \varphi^m_{\bar{c}}[\bar{d}] \}$.
As $X^m_{\bar{c}} \supseteq X^{m+1}_{\bar{c}}$ holds for every $m \geq 0$ and $\bf A$ is a finite structure, there must be an $m^{\bar{c}}$ such that $X^{m^{\bar{c}}}_{\bar{c}} = X^m_{\bar{c}}$ for every $m > m^{\bar{c}}$. Let $m^*$ be the maximum $m^{\bar{c}}$ in $\{m^{\bar{c}} \mid \bar{c} \in A^{\leq |{\cal P}(A \times A)|}\}$. We use $A^{\leq |{\cal P}(A \times A)|}$ to denote the set of tuples of length less than or equal the cardinality of ${\cal P}(A \times A)$. We define the formula $\chi$ as follows:
\begin{equation}
\chi(\bar{x}) \equiv \varphi^{m^*}_{\bar{a}}(\bar{x}) \wedge \bigwedge_{(\bar{a},\bar{c}) \in A^{\leq |{\cal P}(A \times A)|}} \forall \bar{y} (\varphi^{m^*}_{\bar{a}\bar{c}}(\bar{x},\bar{y})  \rightarrow \varphi^{m^*+1}_{\bar{a}\bar{c}}(\bar{x},\bar{y})) \label{isolatingF}
\end{equation}

Finally, we show ${\bf B} \models \chi[\bar{b}] \quad \text{iff} \quad \mathit{tp}_{\bf A}^{\mathrm{FO}_{wo=}}(\bar{a}) = \mathit{tp}_{\bf B}^{\mathrm{FO}_{wo=}}(\bar{b})$. For this we need to check that if $\varphi_{\bar{a}}^{m^*+1}[\bar{b}]$, then $\mathit{tp}_{\bf A}^{\mathrm{FO}_{wo=}}(\bar{a}) = \mathit{tp}_{\bf B}^{\mathrm{FO}_{wo=}}(\bar{b})$. The other direction is immediate. 
\begin{align*}
\text{Let} \; R =& \{ \{(a_1,b_1) \ldots, (a_r, b_r), (a_{r+1}, b_{r+1}), \ldots, (a_l,b_l)\} \mid l \geq r,\\ 
& \qquad \qquad \qquad (a_1, \ldots, a_r, a_{r+1}, \ldots, a_l) \in A^{\leq |{\cal P}(A \times B)|},\\
& \qquad \qquad \qquad (b_1, \ldots, b_r, b_{r+1}, \ldots, b_l) \in B^{\leq |{\cal P}(A \times B)|}\; \text{and}\\
& \qquad \qquad \qquad {\bf B} \models \varphi^{m^*+1}_{a_1 \ldots a_r a_{r+1} \ldots a_l}[b_1, \ldots, b_r, b_{r+1}, \ldots, b_l] \}
\end{align*}

As $\varphi_{\bar{a}}^{m^*+1}[\bar{b}]$, the set $R$ is not empty. 
It follows from~(\ref{equivalence}) that for each $f \in R$, there is a sequence $(I^f_k)_{k \leq m^*+1}$ such that $(I^f_k)_{k \leq m^*+1} : {\bf A} \sim_{m^*+1} {\bf B}$ and $f \in I^f_{m^*+1}$. 

Let $(I_k)_{k \leq m^* + n}$ ($n \geq 1$) be the sequence where $I_k = \bigcup_{f \in R} I^f_k$ for $k \leq m^*+1$ and $I_k = I_{m^*+1}$ for $k > m^*+1$. We claim that, for every $n \geq 1$, it holds that $(I_k)_{k \leq {m^* + n}} : {\bf A} \sim_{m^*+n} {\bf B}$ and $p \in I_{m^*+n}$. We prove it for $n = 2$, the rest then follows immediately. 

By definition we know that $p \in I_{m^*+2}$ and that every $I_k$ is a nonempty set of partial relativeness correspondences. We show that the back and forth conditions hold for $I_{m^*+2}$ and $I_{m^*+1}$. The rest then follows. 

Concerning the forth condition consider any $f \in I_{m^*+2}$ and any $a \in A$. By definition $f \in I_{m^*+1}$ and $p \subseteq f$. As we know that $(I_k)_{k \leq m^*+1}: {\bf A} \sim_{m^*+1} {\bf B}$ holds, there is a $g \in I_{m^*}$ such that $f \subseteq g$ and $a \in \textit{dom}(g)$. Let $g = \{(a_1, b_1,), \ldots, (a_r, b_r), (a_{r+1}, b_{r+1}), \ldots, (a_l, b_l)\}$. Then, by the other direction of~(\ref{equivalence}), ${\bf B} \models \varphi^{m^*}_{a_1 \ldots a_r a_{r+1} \ldots a_l}[b_1, \ldots, b_r, b_{r+1}, \ldots, b_l]$. Therefore, by the implication in~(\ref{isolatingF}) we get ${\bf B} \models \varphi^{m^*+1}_{a_1 \ldots a_r a_{r+1} \ldots a_l}[b_1, \ldots, b_r, b_{r+1}, \ldots, b_l]$ and hence $g \in R$. As $g \in I^g_{m^*+1}$ holds, $g \in I_{m^*+1}$ follows, which proves that the forth condition is met. The same argument can be used to prove that the back condition also holds.

The fact that $(I_k)_{k \leq {m^* + n}} : {\bf A} \sim_{m^*+n} {\bf B}$ and $p \in I_{m^*+n}$ for every $n \geq 1$, together with part~(ii) of Theorem~\ref{FOwo=Characterization}, allow us to conclude that $\mathit{tp}_{\bf A}^{\mathrm{FO}_{wo=}}(\bar{a}) = \mathit{tp}_{\bf B}^{\mathrm{FO}_{wo=}}(\bar{b})$.
\end{proof}

We say that the formula $\chi$ in Lemma~\ref{isolating_formula} {\em isolates} the type $\mathit{tp}_{\bf A}^{\mathrm{FO}_{wo=}}(\bar{a})$. 

Let ${\bf A}$ be a finite structure. It is not difficult to see that if $X^m_{\bar{a}_i} = X^{m+1}_{\bar{a}_i}$ for every $\bar{a}_i \in A^{|{\cal P}(A \times A)|}$, then $X^m_{\bar{a}_i} = X^{m+1}_{\bar{a}_i}$ for every tuple $\bar{a}_i$ of elements from $A$. The relation $\bar{a}_j \in X^{m}_{\bar{a}_i}$ is an equivalence relation on tuples, and the sets $X^{m}_{\bar{a}_i}$ determine a partition of tuples of a given length. Given that there are $|A|^{|{\cal P}(A \times A)|}$ tuples of length $|{\cal P}(A \times A)|$, we can derive the bound $m^* \leq |A|^{|{\cal P}(A \times A)|}$.

Given a formula $\chi$ which isolates the $\mathrm{FO}_{wo=}$-type of a critical tuple $\bar{a}$ in a critical structure ${\bf S}|_W$, we can write an equivalent term $t_\chi$ which evaluates to true in $\bf S$ only for those tuples which have the same $\mathrm{FO}_{wo=}$-type than $\bar{a}$ in ${\bf S}|_W$. 

\begin{lemma}\label{Lemma:IsolatingTerms}

Let $S$ be a state of a reflective parallel algorithm $A$ of vocabulary $\Sigma$, $W$ a bounded exploration witness for $A$, $\bar{a}$ be an $r$-tuple in $(S|_W)^r$, and $\chi$ an isolating formula for the $\mathrm{FO}_{wo=}$-type of $\bar{a}$ in $S|_W$. Then there is a term $t_\chi$ of vocabulary $\Sigma$ such that, for every $\bar{b} \in (V_{{\bf S},W})^r$ we have
\[\mathrm{val}_{S,\mu[\bar{x}\mapsto\bar{b}]}(t_\chi) = \texttt{true} \quad \text{iff} \quad S|_W \models \chi[\bar{b}].\]

\end{lemma}
  
\begin{proof}

For every $\mathrm{FO}_{wo=}$-formula $\varphi$ of vocabulary $\Sigma_W$ we define a corresponding term $t_\varphi$. 

\begin{itemize}

\item If $\varphi(x_1, \ldots, x_r, y)$ is an atomic formula of the form $R_\alpha(x_1, \ldots, x_r, y)$ with $\alpha = \langle (t_1, \ldots, t_r) \mid \psi(y_{1}, \ldots, y_{k}) \rangle$, then $t_\varphi(x_1, \ldots, x_r, y_1, \ldots, y_k)$ is $t_1 = x_1 \wedge \cdots \wedge t_r = x_r \wedge \psi(y_{1}, \ldots, y_{k})$. 

\item If $\varphi$ is a formula of the form $\neg \psi$ or $\psi \wedge \beta$, then $t_\varphi$ is $\neg t_\psi$ or $t_\psi \wedge t_\beta$, respectively.

\item If $\varphi$ is a formula of the form $\exists x (\psi(x))$ or $\forall x (\psi(x))$, then $t_\varphi$ is the term
\begin{gather*}
\langle (z_1, \ldots, z_m) \mid t_{dom}(z_1) \wedge \cdots \wedge t_{dom}(z_m) \wedge t_\psi(z_1, \ldots, z_m) \rangle \neq \oslash \;\text{or} \\
\langle (z_1, \ldots, z_m) \mid \neg ((t_{dom}(z_1) \wedge \cdots \wedge t_{dom}(z_m)) \rightarrow t_\psi(z_1, \ldots, z_m)) \rangle = \oslash \; ,
\end{gather*} 
respectively, where $z_1, \ldots, z_m$ denote the free variable/s in $t_\psi$ that correspond to $x$ and  
\[
t_{dom}(z_i) \equiv \bigvee_{\langle (t_1, \ldots t_r) \mid \psi(x_1, \ldots, x_k)} \in W_S \rangle \begin{pmatrix}\{\!\!\{(x_1, \ldots, x_k, y_1, \ldots, y_r) \mid \\ \psi(x_1, \ldots, x_k) \wedge t_1 = y_1 \wedge \\ \cdots \wedge t_r = y_r  \wedge (z_i = x_1 \vee \\\cdots \vee z_i = x_k \vee z_i = y_1 \vee\\ \cdots \vee z_i = y_r) \}\!\!\}_{z_i} \neq \oslash \end{pmatrix}
\].

\end{itemize}

By induction on $\varphi$ we see that for every tuple $\bar{b}$ of critical elements we get $S|_W \models \varphi[\bar{b}]$ iff $\mathrm{val}_{S,\mu[\bar{x}\mapsto\bar{b}]}(t_\varphi) = \texttt{true}$. Then the isolating formula $\chi$ is just an instance of an $\mathrm{FO}_{wo=}$-formula of vocabulary $\Sigma_W$.
\end{proof}

\subsection{Rules Defined by Update Sets}

For $(f, (a_1, \ldots, a_r), a_0) \in \Delta_A(S)$ let $\chi^{\bar{a}}(x_0, x_1, \ldots, x_r)$ be the isolating formula from Lemma~\ref{isolating_formula} for the $\mathrm{FO}_{wo=}$-type of the critical tuple $\bar{a} = (a_0, a_1, \ldots, a_r)$ in the critical structure $S|_{C_W}$ and let $t^{\bar{a}}_{\chi}(x_0, x_1, \ldots, x_r)$ be its corresponding isolating term from Lemma~\ref{Lemma:IsolatingTerms}. We define $r^S_{A,W}$ as the parallel combination of the update rules
\[ \textbf{forall } x_0, x_1, \ldots, x_r \textbf{ with } t^{\bar{a}}_{\chi}(x_0, x_1, \ldots, x_r) \textbf{ do } f(x_1, \ldots, x_r) := x_0 \]
for all $\bar{a} = (a_0, a_1, \ldots, a_r) \in (S|_{W})^{r+1}$ with $(f, (a_1, \ldots, a_r), a_0) \in \Delta_A(S)$.

\begin{lemma}\label{coro:rule}

If $S$ is a state of the reflective parallel algorithm $A$ and $W$ is a witness set for $A$, then $\Delta(r^S_{A,W}, S) = \Delta_A(S)$.

\end{lemma}  

\begin{proof}

Let $P^S_{A}$ denote the set of \textbf{forall} rules defined above. As $S|_{W}$ is finite and the vocabulary of $S$ has a finite number of function symbols of fixed arity, $P^S_{A}$ is also finite. By Lemma~\ref{lemma:critical_terms} we have $\Delta_A(S) \subseteq \Delta(r^S_{A,W}, S)$. On the other hand, by the semantics of the assignment rule we get $((f, (a_1, \ldots, a_r)), a_0) \in \Delta(r^S_{A,W}, S)$ only if $(a_0, a_1, \ldots, a_r) \in B^r$.  Thus by Lemma~\ref{also_in_update_set} we also have that $\Delta(r^S_{A,W}, S) \subseteq \Delta_A(S)$.
\end{proof}

Note that the rule $r^S_{A,W}$ in Lemma \ref{coro:rule} only involves critical terms that appear in the set $W_S$ determined by the chosen bounded exploration witness $W$ and the extraction function $\beta$. This also implies that the rule is by no means unique.

\subsection{Relative $W$-Similarity}

As in the corresponding proof of the reflective sequential ASM thesis \cite{schewe:scp2022} the following lemmata aim first to extend Lemma \ref{coro:rule} to other states $S^\prime$, i.e. to obtain $\Delta_{r_S}(S^\prime) = \Delta_{\mathcal{A}}(S^\prime)$, and to combine different rules $r_S$ such that the behaviour of $\mathcal{A}$ can be modelled by a combination of such rules on all states.

\begin{lemma}\label{lemma:coincide}

If two states $S$ and $S^\prime$ of $\mathcal{A}$ strongly coincide on $W$, then $\Delta_{r_S}(S^\prime) = \Delta_{\mathcal{A}}(S^\prime)$ holds.

\end{lemma}

\begin{proof}

As $S$ and $S^\prime$ strongly coincide on $W$ we also have $W_S = W_{S^\prime}$, and furthermore, $S$ and $S^\prime$ coincide on $W_S$. As the rule $r_S$ only uses terms in $W_S$, it follows that $\Delta_{r_S}(S) = \Delta_{r_S}(S^\prime)$ holds. Lemma \ref{coro:rule} also states $\Delta_{r_S}(S) = \Delta_{\mathcal{A}}(S)$, and the bounded exploration postulate gives $\Delta_{\mathcal{A}}(S) = \Delta_{\mathcal{A}}(S^\prime)$, which imply $\Delta_{r_S}(S^\prime) = \Delta_{\mathcal{A}}(S^\prime)$ as claimed.
\end{proof}

For the extensions to Lemma \ref{coro:rule} mentioned above we are naturally interested in states $S^\prime$, in which $\mathcal{A}(S^\prime)$ is behaviourally equivalent to $\mathcal{A}(S)$. For this we define that a state $S^\prime$ is {\em $WS$-equivalent} to the state $S$ iff $\beta(\text{val}_{S^\prime}(t)) =  \beta(\text{val}_S(t))$ holds for all $t \in W_{pt}$.

Keeping in mind that the restriction of a state $S$ to $\Sigma_{alg}$ that is to represent a algorithm $\mathcal{A}(S)$ is de facto only relevant for the behaviour on ``standard'' locations, i.e. we are interested in the restrictions $\text{res}(S,\Sigma_S - \Sigma_{alg})$ of states. We therefore use the notation $\Delta^{st}_{\mathcal{A}}(S) = res(\Delta_{\mathcal{A}}(S),\Sigma_S - \Sigma_{alg})$ and similarly $\Delta^{st}_{r_S}(S) = res(\Delta_{r_S}(S),\Sigma_S - \Sigma_{alg})$. Then the following lemma is a straightforward consequence of Lemma \ref{lemma:coincide}. 

\begin{lemma}\label{lem-corollary2}

If the states $S$ and $S^\prime$ are $WS$-equivalent and coincide over $W_S = W_{st} \cup W_\beta$, then we have $\Delta^{st}_{r_S}(S^\prime) = \Delta^{st}_{\mathcal{A}}(S^\prime)$.  

\end{lemma}

\begin{proof}

$WS$-equivalence implies that $W_{S^\prime} = W_S$. As $S$ and $S^\prime$ coincide on $W_S$, they strongly coincide on $W$, which gives $\Delta_{r_S}(S^\prime) = \Delta_{\mathcal{A}}(S^\prime)$ by Lemma \ref{lemma:coincide} and thus the claimed equality of the restricted update sets.
\end{proof}

Now let $\mathcal{C}$ be a class of states. Two states $S_1, S_2 \in \mathcal{C}$ are called {\em $W$-similar relative to $\mathcal{C}$} iff $\sim_{S_1} = \sim_{S_2}$, where the equivalence relation $\sim_{S_i}$ on $W$ is defined by $t \sim_{S_i} t^\prime$ iff $\text{val}_{S_i}(t) = \text{val}_{S_i}(t^\prime)$.

Naturally, we are mainly interested in classes $\mathcal{C}$ that are defined by $WS$-equivalence. We use the notation $[S]$ for the $WS$-equivalence class of the state $S$, i.e. $[S] = \{ S^\prime \mid S^\prime \;\text{is $WS$-equivalent to}\; S \}$. The following two lemmata extend Lemma \ref{coro:rule} to relative $W$-similar states.

\begin{lemma}\label{lem-isomorphism}

If states $S_1, S_2$ are isomorphic, and for state $S$ we have $\Delta^{st}_{r_S}(S_2) = \Delta^{st}_{\mathcal{A}}(S_2)$, then we also get $\Delta^{st}_{r_S}(S_1) = \Delta^{st}_{\mathcal{A}}(S_1)$.

\end{lemma}

\begin{proof}

Let $\sigma$ denote the isomorphism from $S_1$ to $S_2$, i.e. $S_2 = \sigma S_1$. Then $\Delta^{st}_{r_S}(S_2) = \sigma \Delta^{st}_{r_S}(S_1)$ and likewise $\Delta^{st}_{\mathcal{A}}(S_2) = \sigma \Delta^{st}_{\mathcal{A}}(S_1)$. This implies $\sigma \Delta^{st}_{r_S}(S_1) = \sigma \Delta^{st}_{\mathcal{A}}(S_1)$ and further $\Delta^{st}_{r_S}(S_1) = \Delta^{st}_{\mathcal{A}}(S_1)$ by applying $\sigma^{-1}$ to both sides.
\end{proof}

\begin{lemma}\label{lem-relative-equivalence}

If states $S_1$ and $S_2$ are $W$-similar relative to $[S]$, then $\Delta^{st}_{r_{S_1}}(S_2) = \Delta^{st}_{\mathcal{A}}(S_2)$. 

\end{lemma}

\begin{proof}

If we replace every element in the base set of $S_2$ that also belongs to the base set of $S_1$ by a fresh element, we obtain a structure $S^\prime$ isomorphic to $S_2$ and disjoint from $S_1$. By the abstract state postulate $S^\prime$ is a state of $\mathcal{A}$. Furthermore, by construction $S^\prime$ and $S_2$ are also $W$-similar relative to $[S]$. Due to Lemma \ref{lem-isomorphism} it suffices to show $\Delta^{st}_{r_{S_1}}(S^\prime) = \Delta^{st}_{\mathcal{A}}(S^\prime)$, so without loss of generality we may assume that the base sets of $S_1$ and $S_2$ are disjoint. 

Define a new structure $\hat{S}$ by replacing in $S_2$ all critical values $\text{val}_{S_2}(t)$, i.e. $t \in W_S$, by the corresponding value $\text{val}_{S_1}(t)$. As $S_2 \in [S]$ holds, we have $W_{S_2} = W_S$, and due to $W$-similarity relative to $[S]$ we have $\text{val}_{S_1}(t) = \text{val}_{S_1}(t^\prime)$ iff $\text{val}_{S_2}(t) = \text{val}_{S_2}(t^\prime)$ holds, so the structure $\hat{S}$ is well-defined. Furthermore, $\hat{S}$ is isomorphic to $S_2$ and thus a state by the abstract state postulate. We get $\hat{S} \in [S]$ by construction. Furthermore, $S_1$ and $\hat{S}$ coincide on $W_S$, so Lemma \ref{lem-corollary2} implies that $\Delta^{st}_{r_{S_1}}(\hat{S}) = \Delta^{st}_{\mathcal{A}}(\hat{S})$ holds. Using again Lemma \ref{lem-isomorphism} completes the proof.
\end{proof}

Next we define a rule $r_{[S]}$ for a whole $WS$-equivalence class $[S]$. For $S_1 \in [S]$ let $\varphi_{S_1}$ be the following Boolean term:
\[ \bigwedge_{\substack{t_i, t_j \in W_{st} \,\cup\, W_{\beta} \\ \text{val}_{S_1}(t_i) = \text{val}_{S_1}(t_j)}} t_i = t_j \quad \wedge \bigwedge_{\substack{t_i, t_j \in W_{st} \,\cup\, W_{\beta} \\ \text{val}_{S_1}(t_i) \neq \text{val}_{S_1}(t_j)}} \neg (t_i = t_j). \] 

Clearly, a state $S_2 \in [S]$ satisfies $\varphi_{S_1}$ iff $S_1$ and $S_2$ are $W$-similar relative to $[S]$. As $W$ is finite, we obtain a partition of $[S]$ into classes $[S]_1 ,\dots, [S]_n$ such that two states belong to the same class $[S]_i$ iff they are $W$-similar relative to $[S]$. We choose representatives $S_1 ,\dots, S_n$ for these classes and define the rule $r_{[S]}$ by
\[ \texttt{PAR}\; (\texttt{IF}\; \varphi_{S_1} \;\texttt{THEN}\; r_{S_1} \;\texttt{ENDIF}) \; \dots \;
(\texttt{IF}\; \varphi_{S_n} \;\texttt{THEN}\; r_{S_n} \;\texttt{ENDIF}) \; \texttt{ENDPAR} \]

Using this rule we obtain the following lemma, which is a straightforward consequence of the previous lemmata. 

\begin{lemma}\label{lem-relatBehavEquiv}  

For every state $S$ and every state $S^\prime \in [S]$ we have $\Delta^{st}_{r_{[S]}}(S^\prime) = \Delta^{st}_{\mathcal{A}}(S^\prime)$.

\end{lemma} 

\begin{proof}

There is exactly one class $[S]_i$ with representing state $S_i$ such that $S^\prime \in [S]_i$ holds. Then $\text{val}_{S^\prime}(\varphi_j)$ is \textbf{true} iff $j = i$. Then we get $\Delta^{st}_{r_{[S]}}(S^\prime) = \Delta^{st}_{r_{S_i}}(S^\prime) = \Delta^{st}_{\mathcal{A}}(S^\prime)$ using the definition of $r_{[S]}$ and Lemma \ref{lem-relative-equivalence}.
\end{proof}

\subsection{Differences of Trees}

We know that for every state $S$ there is a well-defined, consistent update set $\Delta(S)$ such that $S^\prime = S+\Delta(S)$ is the successor state of $S$. Actually, $\Delta(S)$ arises as the difference between $S$ and $S^\prime$. If the states contain locations with bulk values assigned to them, then it becomes also important to provide means for the expression of the difference of such values. Here we concentrate only on the tree values assigned to \textit{pgm\/}. Let $T_L$ be the set of trees with labels in $L$ and values in a universe $\mathcal{U}$ as used in the definition of backgrounds.

\begin{lemma}\label{prop-tree-difference}

If $t, t^\prime \in T_L$ are trees that represent ASMs in successive states, then there is a tree algebra term $\theta$ such that $t^\prime = \theta(t)$.

\end{lemma}

\begin{proof}

First consider the subtrees $t_\mathit{sig}$ and $t_\mathit{sig}^\prime$ representing the signatures in $S$ and $\tau(S)$, respectively. As we assume that only new function symbols are added, we obtain immediately that $t_\mathit{sig}^\prime$ can be obtained from $t_\mathit{sig}$ by simply applying an algebra term of the form 
\[ \textit{right\_extend\/}(t_\mathit{sig}, \textit{label\_hedge\/}(\texttt{func}, \langle f_1 \rangle \, \langle a_1 \rangle) \dots \textit{label\_hedge\/}(\texttt{func}, \langle f_k \rangle \, \langle a_k \rangle)).\] 

Concerning the subtrees $t_\mathit{rule}$ and $t_\mathit{rule}^\prime$ representing the rules in $S$ and $\tau(S)$, respectively, we proceed by building the algebra expression $\theta$ by applying the following recursive rules while traversing  $t_\mathit{rule}^\prime$ in reverse breadth first order, i.e.\@ from the leaves of $t_\mathit{rule}^\prime$ to its root.

\begin{enumerate}

\item If $o$ is a node in $t_\mathit{rule}^\prime$ representing an assignment rule that does not appear in $t_\mathit{rule}$, then $\textit{subtree\/}(o)$ takes the form
\[\textit{label\_hedge}(\texttt{update},\texttt{func} \langle f \rangle \texttt{term} \langle t_1 \dots t_n \rangle \texttt{term} \langle t_0 \rangle) . \]

\item If $o$ is a node in $t_\mathit{rule}^\prime$ representing a partial assignment rule that does not appear in $t_\mathit{rule}$, then $\textit{subtree\/}(o)$ takes the form
\[\textit{label\_hedge}(\texttt{partial},\texttt{func} \langle f \rangle \texttt{func} \langle op \rangle \texttt{term} \langle t_1 \dots t_n \rangle \texttt{term} \langle t_1^\prime \dots t_m^\prime \rangle) . \]

\item If $o$ is a node in $t_\mathit{rule}^\prime$ such that $\textit{subtree\/}(o)$ occurs as a subtree of $t_\mathit{rule}$ and $o'$ is the root of this subtree in $t_\mathit{rule}$, then $\textit{subtree\/}(o)$ simply takes the form $\textit{subtree\/}(o')$.  

\item If $o$ is a node in $t_{rule}^\prime$ with children $o_1 ,\dots, o_l$ and we can write $\textit{subtree\/}(o_i) = \theta_i(t)$, then we obtain 
\[ \textit{subtree\/}(o) = \textit{label\_hedge}(a, \theta_1(t) \dots \theta_l(t)) \]
with a label $a \in \{ \texttt{if}, \texttt{par}, \texttt{forall}, \texttt{let}, \texttt{import} \}$.

\end{enumerate}
\end{proof}

\begin{lemma}\label{prop-tree-update}

For trees $t, t^\prime \in T_L$ that represent parallel ASMs in successive states, there exists a compatible update multiset $\ddot{\Delta}$ defined by updates and shared updates on the nodes of $t$ such that $\ddot{\Delta}$ collapses to $\Delta = \{ (\textit{root\/}(t), t^\prime) \}$.

\end{lemma}

\begin{proof}

Let $o^\prime$ be the unique successor node of the root of $t^\prime$ labelled by \texttt{signature}.
For the subtree of $t'$ representing the signature we then obtain a rule of the form
\[ o^\prime := \textit{right\_extend\/}(t_{sig}, \textit{label\_hedge\/}(\texttt{func}, \langle f_1 \rangle \, \langle a_1 \rangle) \dots \textit{label\_hedge\/}(\texttt{func}, \langle f_k \rangle \, \langle a_k \rangle)) , \]
where the subtree $t_{sig}$ represents the signature in the current state. 

For the subtree representing the rule we simply define rules for the four cases used in the proof of Lemma \ref{prop-tree-difference}:

\begin{enumerate}

\item If $o$ is a node in $t_{rule}^\prime$ representing an assignment rule that does not appear in $t$, the rule takes the form
\[ o := \textit{label\_hedge}(\texttt{update},\texttt{func} \langle f \rangle \texttt{term} \langle t_1 \dots t_n \rangle \texttt{term} \langle t_0 \rangle) . \]

\item If $o$ is a node in $t_{rule}^\prime$ representing a partial assignment rule that does not appear in $t$, then the rule takes the form
\[ o := \textit{label\_hedge}(\texttt{partial},\texttt{func} \langle f \rangle \texttt{func} \langle op \rangle \texttt{term} \langle t_1 \dots t_n \rangle \texttt{term} \langle t_1^\prime \dots t_m^\prime \rangle) . \]

\item In case $o$ is a node in $t_\mathit{rule}^\prime$ such that $\textit{subtree\/}(o)$ occurs as a subtree of $t_\mathit{rule}$ and $o'$ is the root of this subtree in $t_\mathit{rule}$, then the rule takes the form 
\begin{gather*}
\texttt{LET}\; o^\prime = \textbf{I} o_{kl} . \textit{root\/}(t) \prec_c^k o_{kl} \wedge \exists o_1,\dots,o_l . o_1 \prec_s o_2 \prec_s \dots \prec_s o_l \prec_s o_{kl} \wedge \\
\neg \exists o_0 . o_0 \prec_s o_1 \;\texttt{IN}\; o := \textit{subtree\/}(o^\prime)
\end{gather*}

\item For all other nodes $o$ the rule takes the form $o := \textit{label\_hedge}(a, t_1 \dots t_l)$ with a label $a \in \{ \texttt{if}, \texttt{par}, \texttt{forall}, \texttt{let}, \texttt{import} \}$ and terms $t_1 ,\dots, t_l$ that are used in the assignment rules for the children of $o$.

\end{enumerate}

Each of these rules for all nodes of $t^\prime$ defines a partial update rule for the location \textit{pgm\/}, which defines the sought update multiset $\ddot{\Delta}$ and the update set $\Delta$.
\end{proof}

\subsection{Tree Updates}

We need to combine the partial results obtained so far into the construction of a rASM $\mathcal{M}$ that is behaviourally equivalent to the given RA $\mathcal{A}$. Lemma \ref{coro:rule} shows that each state transition can be expressed by a rule, i.e. for each state $S$ we have a rule $r_S$ with $\Delta_{r_S}(S) = \Delta_{\mathcal{A}}(S)$, hence $\tau(S) = S + \Delta_{r_S}(S)$. Lemma \ref{lem-relatBehavEquiv} gives a single rule $r_{[S]}$ for each class $[S]$ with $\Delta^{st}_{r_{[S]}}(S^\prime) = \Delta^{st}_{\mathcal{A}}(S^\prime)$ for all $S^\prime \in [S]$, hence $res(\tau(S^\prime), \Sigma_{\tau(S^\prime)} - \Sigma_{alg}) = res(S^\prime + \Delta_{r_{[S]}}(S^\prime), \Sigma_{\tau(S^\prime)} - \Sigma_{alg})$. The latter result already captures the updates on the ``standard'' part of the state by an ASM rule, but the whole rule $r_{[S]}$ updates also locations with function symbols in $\Sigma_{alg}$ instead of \textit{pgm\/}.

What we need is a different representation of a rule in a state extending the restriction to $\Sigma_S - \Sigma_{alg}$ and using a location \textit{pgm\/}. For any state $S$ of the RA $\mathcal{A}$ let $\Phi(S)$ denote a structure over $(\Sigma_S - \Sigma_{alg}) \cup \{ \textit{pgm\/} \}$, in which the location \textit{pgm\/} captures this intended representation. The structure $\Phi(S)$ should be built as follows:

\begin{enumerate}

\item We must have $res(\Phi(S), \Sigma_S - \Sigma_{alg}) = res(S, \Sigma_S - \Sigma_{alg})$.

\item The value $\text{val}_S(\textit{pgm})$ must be a tree composing a signature subtree and a rule subtree, i.e. $\text{val}_S(\textit{pgm}) =$
\[ \textit{label\_hedge\/}(\texttt{pgm}, \textit{label\_hedge\/}(\texttt{signature}, s_0 s_1 \dots s_k), \textit{label\_hedge\/}(\texttt{rule}, r) . \]

\item The signature subtree simply lists the signature $\Sigma_{\Phi(S)} = (\Sigma_S - \Sigma_{alg}) \cup \{ \textit{pgm\/} \}$, so with $f_0 = \textit{pgm\/}$, $\{ f_1 ,\dots, f_k \} = \Sigma_S - \Sigma_{alg}$, $n_0 = 0$, and $ar(f_i) = n_i$ for all $i$ each of the subtrees $s_i$ takes the form
\[ \textit{label\_hedge\/}(\texttt{func}, \texttt{name} \langle f_i \rangle \; \texttt{arity} \langle n_i \rangle) . \]

\item The rule subtree captures a representation of a rule $r_{\Phi(S)}$, which must be a parallel composition of two parts: (a) a rule capturing the updates to $res(\Phi(S), \Sigma_S - \Sigma_{alg})$, (b) a rule capturing updates to \textit{pgm\/}.

\end{enumerate}

Concerning the updates to $res(\Phi(S), \Sigma_S - \Sigma_{alg})$ we know from Lemma \ref{lem-relatBehavEquiv} that these are captured by the rule $r_{[S]}$ restricted to $\Sigma_S - \Sigma_{alg}$, so the representation as a tree is done in the standard way. Let $t_{r_{[S]}}$ denote this tree.

The remaining problem is to find a rule capturing the updates to the tree assigned to \textit{pgm\/} in any state $\Phi(S)$. For this we will proceed analogously to Lemmata \ref{coro:rule}-\ref{lem-relatBehavEquiv} focusing on finitely many changes to trees. In particular, it will turn out that a fixed tree representing a rule $r^{self}$ will be sufficient for all states $S$.

We start with a tree $\hat{t}_S$ satisfying the requirements from (2), (3) and (4)(a) above, but ignoring updates to \textit{pgm\/}. This tree is defined as 
\[ \textit{label\_hedge\/}(\texttt{pgm}, \textit{label\_hedge\/}(\texttt{signature}, s_0 s_1 \dots s_k), \textit{label\_hedge\/}(\texttt{rule}, t_{r_{[S]}}) . \]

Furthermore, we let $\hat{\Phi}(S)$ denote the structure defined by $res(\hat{\Phi}(S), \Sigma_S - \Sigma_{alg}) = res(S, \Sigma_S - \Sigma_{alg})$ and $\text{val}_{\hat{\Phi}(S)}(\textit{pgm\/}) = \hat{t}_S$.

\begin{lemma}\label{lem-tree-update}

There exists a rule $r_S^{self}$ with $\Delta_{r_S^{self}}(\hat{\Phi}(S)) = \{ (\textit{pgm\/}, \hat{t}_{\tau(S)} ) \}$. The rule uses only the operators of the tree algebra and terms of the form $\textit{raise\/}(t)$ with $t$ appearing in $\text{val}_{\hat{\Phi}(S)}(\textit{pgm\/})$.

\end{lemma}

\begin{proof}

Concerning the signature finitely many new function symbols are added to $\Sigma_S$ by the transition of $\mathcal{A}$ to $\tau(S)$. So we can define a rule $r_{\hat{\Phi}(S)}^{sig}$ as a finite parallel composition of rules $\textit{NewFunction\/}(f_i,n_i)$ ($i = 1,\dots,k$), where $\textit{NewFunction\/}(f,n)$ is defined as
\begin{gather*}
\texttt{LET}\; \textit{sign\/}\; = \textbf{I} o . ( \textit{root\/}(\textit{pgm\/}) \prec_c o \wedge \textit{label\/}(o) = \texttt{signature} ) \; \texttt{IN} \\
\textit{sign\/}\; \leftleftarrows^{\textit{right\_extend\/}}\; \textit{label\_hedge\/}(\texttt{func}, \langle f \rangle \; \langle n \rangle ) 
\end{gather*}

According to Lemma \ref{coro:rule} we have $\tau(S) = S + \Delta_{r_S}(S)$, so every new function symbol $f_i$ must be taken from the reserve $\Sigma_{res}$ and its arity $n_i$ must be the value of a term used in $r_S$. Such terms also occur in $r_{[S]}$, hence they must have the form $\textit{raise\/}(t)$ with $t$ appearing in $\text{val}_{\hat{\Phi}(S)}(\textit{pgm\/})$.

Concerning the subtrees representing the rule represented in \textit{pgm\/} the rules $r_{[S]}$ and $r_{[\tau(S)]}$ take the form
\[ \texttt{PAR}\; \texttt{IF}\; \varphi_1 \;\texttt{THEN}\; r_1 \;\texttt{ENDIF}\; \dots \texttt{IF}\; \varphi_l \;\texttt{THEN}\; r_l \;\texttt{ENDIF} \; \texttt{ENDPAR}\; , \]
in which the rules $r_i$ are parallel compositions of \texttt{forall}-rules defined by isolating formulae and exploiting either assignment rules $f(t_1,\dots,t_n) := t_0$ with $f \in \Sigma_S - \Sigma_{alg}$ and partial update rules $f(t_1,\dots,t_n) \leftleftarrows^{op} t_1^\prime ,\dots, t_m^\prime$. These are represented by trees of the form
\[ \textit{label\_hedge}(\texttt{update},\texttt{func} \langle f \rangle \; \texttt{term} \langle t_1 \dots t_n \rangle \; \texttt{term} \langle t_0 \rangle) \]
and
\[ \textit{label\_hedge}(\texttt{partial},\texttt{func} \langle f \rangle \; \texttt{func} \langle op \rangle \; \texttt{term} \langle t_1 \dots t_n \rangle \; \texttt{term} \langle t_1^\prime \dots t_m^\prime \rangle) , \]

respectively. According to Lemma \ref{prop-tree-difference} $t_{r_{[\tau(S)]}}$ can be defined by a tree algebra expression on $t_{r_{[S]}}$ with values of the form $\textit{drop\/}(t)$ for $t$ appearing in $r_{[S]}$. Again, these terms must have the form $\textit{raise\/}(t)$ with $t$ appearing in $\text{val}_{\hat{\Phi}(S)}(\textit{pgm\/})$. Lemma \ref{prop-tree-update} which defines an update set $\Delta$, which using all proof arguments for Lemma \ref{coro:rule} defines a rule $r_{\hat{\Phi}(S)}^{rule}$, and $r_S^{self}$ is then the parallel composition of $r_{\hat{\Phi}(S)}^{sig}$ and $r_{\hat{\Phi}(S)}^{rule}$.
\end{proof}

Lemma \ref{lem-tree-update} shows that we can always find a rule that transforms one tree representation into another one. If we had chosen a different self-representation, we would need similarly powerful manipulation operators that ensure an analogous result. For our proof here it will be essential to show that only finitely many such rules are needed. We will show this by the following sequence of lemmata, by means of which we extend the applicability of the rule $r_S^{\textit{pgm\/}}$ to trees $\hat{\Phi}(S^\prime)$ for other states $S^\prime$.

\begin{lemma}\label{lem-tree-coincidence}

If states $S$ and $S^\prime$ coincide on $W$, then we have $\Delta_{r_S^{self}}(\hat{\Phi}(S^\prime)) = \{ (\textit{pgm\/}, \hat{t}_{\tau(S^\prime)} ) \}$.

\end{lemma}

\begin{proof}

The updates from $\mathcal{A}(S)$ to $\mathcal{A}(\tau(S))$ are defined by $\text{res}(\Delta_{\mathcal{A}}(S), \Sigma_{alg})$, so we have $\text{res}(S, \Sigma_{alg}) + \text{res}(\Delta_{\mathcal{A}}(S), \Sigma_{alg}) = \text{res}(\tau(S), \Sigma_{alg})$. Analogously, the changes from $\mathcal{A}(S^\prime)$ to $\mathcal{A}(\tau(S^\prime))$ are defined by $\text{res}(\Delta_{\mathcal{A}}(S^\prime), \Sigma_{alg})$: we have $\text{res}(S^\prime, \Sigma_{alg}) + \text{res}(\Delta_{\mathcal{A}}(S^\prime), \Sigma_{alg}) = \text{res}(\tau(S^\prime), \Sigma_{alg})$. Then this also applies to the restrictions of $\mathcal{A}(S)$ and $\mathcal{A}(S^\prime)$ to the states defined over $\Sigma_S - \Sigma_{alg}$ and $\Sigma_{S^\prime} - \Sigma_{alg}$, respectively. These restricted algorithms are behaviourally equivalent to $r_{[S]}$ and $r_{[\tau(S)]}$.

Thus, the updates from $r_{[S^\prime]}$ to $r_{[\tau(S^\prime)]}$ are defined by $\text{res}(\Delta_{\mathcal{A}}(S^\prime), \Sigma_{alg})$. As $S$ and $S^\prime$ coincide on $W$ the bounded exploration postulate implies that $\text{res}(\Delta_{\mathcal{A}}(S^\prime), \Sigma_{alg}) = \text{res}(\Delta_{\mathcal{A}}(S), \Sigma_{alg})$ holds. As the updates from $r_{[S]}$ to $r_{[\tau(S)]}$ using their tree representations are equivalently expressed by $r_S^{\textit{self\/}}$, Lemma \ref{lem-tree-update} implies that the updates from the tree representation of $r_{[S^\prime]}$ to the tree representation of $r_{[\tau(S^\prime)]}$ are also defined by $r_S^{\textit{self\/}}$ as claimed.
\end{proof}

\begin{lemma}\label{lem-tree-isomorphism}

If states $S_1, S_2$ are isomorphic, and for a state $S$ we have $\Delta_{r_S^{\textit{self\/}}}(\hat{\Phi}(S_2)) = \{ (\textit{pgm\/}, \hat{t}_{S_2} ) \}$, then we also get $\Delta_{r_S^{\textit{self\/}}}(\hat{\Phi}(S_1)) = \{ (\textit{pgm\/}, \hat{t}_{S_1} ) \}$.

\end{lemma}

\begin{proof}

Let $\sigma$ denote the isomorphism from $S_1$ to $S_2$, i.e. $S_2 = \sigma S_1$. Then $\Delta_{r_S^{\textit{self\/}}}(\hat{\Phi}(S_2)) = \sigma \Delta_{r_S^{\textit{self\/}}}(\hat{\Phi}(S_1))$ and likewise $\hat{t}_{S_2} = \sigma \hat{t}_{S_1}$. 

This implies $\sigma \Delta_{r_S^{\textit{self\/}}}(\hat{\Phi}(S_1)) = \sigma \{ (\textit{pgm\/}, \hat{t}_{S_1}) \}$ and further $\Delta_{r_S^{\textit{self\/}}}(\hat{\Phi}(S_1)) = \{ (\textit{pgm\/}, \hat{t}_{S_1}) \}$ by applying $\sigma^{-1}$ to both sides.
\end{proof}

\subsection{$W$-Similarity}

For Lemma \ref{lem-relative-equivalence} we exploited relative $W$-similarity, which exploits terms in classes $[S]$ defined by the evaluation of the bounded exploration witness $W$. As the class $[S]$ depends on the evaluation of the terms in $W_{pt}$, also relative equivalence implicitly depends on $W_{pt}$. Now we require a notion of $W$-similarity that is grounded only on $W$. Two states $S_1, S_2$ are called {\em $W$-similar} iff $\sim_{S_1} = \sim_{S_2}$ holds, where the equivalence relation $\sim_{S_i}$ on $W$ is defined by $t \sim_{S_i} t^\prime$ iff $\text{val}_{S_i}(t) = \text{val}_{S_i}(t^\prime)$.

Then the proof of the following lemma will be quite analogous to the proof of Lemma \ref{lem-relative-equivalence}.

\begin{lemma}\label{lem-tree-similarity}

\ If states $S$ and $S^\prime$ are $W$-similar, then we have $\Delta_{r_S^{self}}(\hat{\Phi}(S^\prime)) = \{ (\textit{pgm\/}, \hat{t}_{\tau(S^\prime)} ) \}$.

\end{lemma}

\begin{proof}

If we replace every element in the base set of $S^\prime$ that also belongs to the base set of $S$ by a fresh element, we obtain a structure $S^{\prime\prime}$ isomorphic to $S^\prime$ and disjoint from $S$. By the abstract state postulate $S^{\prime\prime}$ is also a state of $\mathcal{A}$. Furthermore, by construction $S^\prime$ and $S^{\prime\prime}$ are also $W$-similar to $S$. Lemma \ref{lem-isomorphism} already covers equality on the ``standard part'' of the signature, so it suffices to show $\Delta_{r_S^{self}}(\hat{\Phi}(S^{\prime\prime})) = \{ (\textit{pgm\/}, \hat{t}_{\tau(S^{\prime\prime})} ) \}$, so without loss of generality we may assume that the base sets of $S$ and $S^\prime$ are disjoint. 

Define a new structure $\hat{S}$ by replacing in $S^\prime$ all values $\text{val}_{S^\prime}(t)$ with a critical term $t \in W$ by the corresponding value $\text{val}_{S}(t)$. Due to $W$-similarity we have $\text{val}_{S}(t) = \text{val}_{S}(t^\prime)$ iff $\text{val}_{S^\prime}(t) = \text{val}_{S^\prime}(t^\prime)$ holds, so the structure $\hat{S}$ is well-defined. Furthermore, $\hat{S}$ is isomorphic to $S^\prime$ and thus a state by the abstract state postulate. Furthermore, $S$ and $\hat{S}$ coincide on $W$, so Lemma \ref{lem-tree-coincidence} implies that $\Delta_{r_S^{self}}(\hat{\Phi}(\hat{S})) = \{ (\textit{pgm\/}, \hat{t}_{\tau(\hat{S})} ) \}$ holds. Using again Lemma \ref{lem-tree-isomorphism} completes the proof.
\end{proof}

Using Lemma \ref{lem-tree-similarity} we can exploit that there are only finitely many $W$-similarity classes to define a single rule $r^{\textit{self\/}}$ for all states. Let $\varphi_S$ be the following Boolean term:
\[ \bigwedge_{\substack{t_i, t_j \in W \\ \text{val}_{S}(t_i) = \text{val}_{S}(t_j)}} t_i = t_j \quad \wedge \bigwedge_{\substack{t_i, t_j \in W \\ \text{val}_{S}(t_i) \neq \text{val}_{S}(t_j)}} \neg (t_i = t_j). \] 

Clearly, a state $S^\prime$ satisfies $\varphi_S$ iff $S$ and $S^\prime$ are $W$-similar. As $W$ is finite, we obtain a partition of $\mathcal{S}$ into classes $[S_1] ,\dots, [S_n]$ with representatives $S_i$ ($i=1,\dots,n$) such that a state $S$ belongs to the class $[S_i]$ iff $S$ and $S_i$ are $W$-similar. We define the rule $r^{\textit{self\/}}$ by
\[ \texttt{PAR}\; (\texttt{IF}\; \varphi_{S_1} \;\texttt{THEN}\; r_{S_1}^{\textit{self\/}} \;\texttt{ENDIF}) \; \dots \;
(\texttt{IF}\; \varphi_{S_n} \;\texttt{THEN}\; r_{S_n}^{\textit{self\/}} \;\texttt{ENDIF}) \; \texttt{ENDPAR} \]

Using this rule we obtain the following lemma, which is a straightforward consequence of the previous lemmata. 

\begin{lemma}\label{lem-treeBehavEquiv}  

For all states $S$ we have $\Delta_{r^{self}}(\hat{\Phi}(S)) = \{ (\textit{pgm\/}, \hat{t}_{\tau(S)} ) \}$.

\end{lemma} 

\begin{proof}

There is exactly one class $[S_i]$ with representing state $S_i$ such that $S \in [S_i]$ holds. Then $\text{val}_{S}(\varphi_j)$ is \textbf{true} iff $j = i$. Then we get $\Delta_{r^{self}}(\hat{\Phi}(S)) = \Delta_{r_{S_i}^{self}}(\hat{\Phi}(S)) = \{ (\textit{pgm\/}, \hat{t}_{\tau(S)} ) \}$ using the definition of $r^{\textit{self\/}}$ and Lemma \ref{lem-tree-similarity}.
\end{proof}

We now extend the tree $\hat{t}_S$ to a definition of a tree $t_S$ satisfying all the requirements from (2), (3) and (4) above including now the updates to \textit{pgm\/}. This tree is defined as 
\begin{gather*}
\textit{label\_hedge\/}(\texttt{pgm}, \textit{label\_hedge\/}(\texttt{signature}, s_0 s_1 \dots s_k), \\ \textit{label\_hedge\/}(\texttt{rule}, \textit{label\_hedge\/}(\texttt{par}, t_{r_{[S]}} \; t_{r^{\textit{self\/}}})) , 
\end{gather*}

where $t_{r^{\textit{self\/}}}$ is the tree representation of the rule $r^{\textit{self\/}}$. Furthermore, we let $\Phi(S)$ denote the structure defined by $res(\Phi(S), \Sigma_S - \Sigma_{alg}) = res(S, \Sigma_S - \Sigma_{alg})$ and $\text{val}_{\Phi(S)}(\textit{pgm\/}) = t_S$. We use this to define an rASM $\mathcal{M}$.

\begin{lemma}\label{lem-rsasm}

For each RA $\mathcal{A}$ we obtain an rASM $\mathcal{M}$ with the set $\mathcal{S}_{\mathcal{M}} = \{ \Phi(S) \mid S \in \mathcal{S} \}$ of states, the set $\mathcal{I}_{\mathcal{M}} = \{ \Phi(S) \mid S \in \mathcal{I} \}$ of initial states, and the state transition function $\tau_{\mathcal{M}}$ with $\tau_{\mathcal{M}}(\Phi(S)) = \Phi(\tau(S))$.

\end{lemma}

\begin{proof}

According to the abstract state postulate all initial states $S_0, S_0^\prime \in \mathcal{I}$ are defined over the same signature $\Sigma_0 \cup \Sigma_{alg}$. As $\mathcal{I}$ is closed under isomorphisms, this also holds for $\mathcal{I}_{\mathcal{M}}$. As $\mathcal{A}(S_0) = \mathcal{A}(S_0^\prime)$ holds, the states $\Phi(S_0)$ and $\Phi(S_0^\prime)$ coincide on \textit{pgm\/}.

We have $\textit{raise\/}(\textit{rule\/}(\text{val}_{\Phi(S)}(\textit{pgm\/}))) = \texttt{PAR}\; r_{[S]} \; r^{\textit{self\/}} \; \texttt{ENDPAR}$. Denoting this rule as $r_{\Phi(S)}$ we obtain
\[ \Delta_{r_{\Phi(S)}}(\Phi(S)) = \Delta^{st}_{r_{[S]}}(S) \cup \{ (\textit{pgm\/}, t_{\tau(S)} ) \} , \]

which implies $\Phi(S) + \Delta_{r_{\Phi(S)}}(\Phi(S)) = \Phi(\tau(S))$ using Lemmata \ref{lem-relatBehavEquiv} and \ref{lem-treeBehavEquiv}.
\end{proof}

Finally, we combine all our lemmata into a proof of Theorem \ref{thm:characterisation}.

\begin{proof}[Proof of Theorem \ref{thm:characterisation}]

Given the RA $\mathcal{A}$ we define the rASM $\mathcal{M}$ as in Lemma \ref{lem-rsasm}. It remains to show that $\mathcal{A}$ and $\mathcal{M}$ are behaviourally equivalent. For this let $S_0, S_1, \dots$ be an arbitrary run of $\mathcal{A}$. Then according to Lemma \ref{lem-rsasm} $\Phi(S_0), \Phi(S_1), \dots$ is a run of $\mathcal{M}$, which defines the required bijection between runs of $\mathcal{A}$ and $\mathcal{M}$. 

Due to our construction of the states $\Phi(S)$ we further have $\textit{res\/}(\mathcal{A}, \Sigma_{S} - \Sigma_{alg}) = \textit{res\/}(\mathcal{M}, \Sigma_{S} - \Sigma_{alg})$, which implies the first property required for behavioural equivalence.

The restriction of the algorithm $\mathcal{A}(S)$ to $\Sigma_{S} - \Sigma_{alg}$ is expressed by the rule $r_{[S]}$. The algorithm represented in $\Phi(S)$ is expressed by the rule 
\[ \texttt{PAR}\; r_{[S]} \; r^{\textit{self\/}} \;\texttt{ENDPAR} \]
and its restriction to $\Sigma_{S} - \Sigma_{alg}$ is also expressed by the rule $r_{[S]}$, which implies the second property required for behavioural equivalence.
\end{proof}

\section{Related Work}\label{sec:others}

This article continues the investigation of behavioural theories for reflective algorithms extending our previous work on reflective sequential algorithms \cite{schewe:scp2022}. Linguistic reflection as such has a long tradition in programming languages, and behavioural theories have by now been established as an important tool for studying expressiveness of rigorous methods.

\subsection{Linguistic Reflection}

Adaptivity (or self-adaptation) refers to the ability of a system to change its own behaviour. In the context of programming this concept, known under the term {\em linguistic reflection}, appears already in the 1950s in LISP~\cite{smith:popl1984}, where programs and data are both represented uniformly as lists, and thus programs represented as data can be executed dynamically by means of an evaluation operator. This has been preserved in many functional programming languages such as SCHEME \cite{sussman:1998}, HASKELL \cite{peyton:2003} and others. 

While the concept of linguistic reflection is old, it first disappeared from many programming languages, but survived in niches, e.g. in the language POP~\cite{popplestone:fac2002}. The reason is that it is difficult to maintain control of the desired behaviour of a program when this behaviour is subject to on-the-fly changes. Reflection resurfaced in the context of persistent programming and object stores, where strong typing was perceived as a straightjacket. Stemple and Sheard were the first to point to the limitations of expressiveness of parametric polymorphism showing that the common natural join cannot be expressed that way, but it could easily be captured in their reflective language AdabtPL \cite{stemple:edbt1990}. Based on this observation they promoted linguistic reflection as a necessary tool in persistent programming emphasising that it appears naturally in connection with operations on bulk data types \cite{stemple:dbpl1991}, provides a natural tool to control inheritance \cite{sheard:iccl1988,sheard:cola1992}, and supports the verification of transaction safety \cite{sheard:tods1989}.

Tools to support reflection such as START \cite{kirby:pos1994} and Chord \cite{tauber:saso2010} as well as reflection middleware \cite{dearle:corr2010} were created; the emphasis was mostly on meta-level programming to compute types, by means of which the restrictions of strongly typed languages were mitigated without giving up strong typing as a means of quality control. The Ph.D. thesis of Kirby provides a good survey on the many achievements concerning reflection in persistent programming and object stores \cite{kirby:1992}. It was shown how reflection can enable higher-level genericity in object-oriented databases \cite{schewe:comad1994}. Also a reflective extension of Java \cite{kirby:spe1998} has been developed. The concept of run-time linguistic reflection as in \cite{cooper:pos1994} usually involves a callable compiler, whereas compile-time linguistic reflection expands macros as part of the compilation process. This has been generalised for reflective programming languages by Stemple and others \cite{stemple:2000}. The expressiveness of linguistic reflection in database programming has been investigated by Van den Bussche and others \cite{bussche:jcss1996}.

With an upcoming interest in general self-adaptive systems \cite{balasubramaniam:woss2004} reflection gained interest outside the narrow field of persistent database programming languages. Sheard emphasised that reflection should be considered a natural component of future programming languages \cite{sheard:sigplan2004}, and he promoted TRPL (type-safe reflective programming language) as a prototype for this direction. As reflective programming can always be seen as computation on two levels, the object level where states are updated and the meta-level where the program itself is updated, meta programming was introduced \cite{sheard:saig2001} and among others demonstrated by an extension of the functional language ML \cite{taha:tcs2000}. In this context it was shown on a rather practical level how meta programming supports adaptivity \cite{harrison:saig2001}, and how it can be exploited for refinement \cite{greenwood:ewspt2001}. Despite these achievements, it became quiet again for a while around reflection. 

Adaptivity in computing systems has another source in natural computing, an area where algorithms are designed in a way to mimick processes in nature. A common motivation is the capture of complex adaptive systems \cite{holland:jssc2006}. This area developed quite independently from the work above on linguistic reflection. The most common representatives of this research direction are genetic programming \cite{holland:1992,holland:scholar2012,goldberg:1989,koza:1993}, DNA computing \cite{paun:1998} and membrane computing \cite{paun:2002}. Same as for reflection the research on logical foundations of natural computing also dates back to the early days of programming languages \cite{holland:jacm1962}. The usefulness of genetic algorithms has been shown by applications in statistical machine learning \cite{goldberg:ml1988}, in particular for classification problems \cite{booker:ai1989}, for general heuristic search by simulated annealing \cite{mahfoud:parcom1995}, and also for circuit synthesis \cite{koza:softc2004}. More recently, this has been generalised to a theory of evolving reaction systems in general \cite{ehrenfeucht:tcs2017}. Recently, adaptivity has become a key aspect of cyber-physical systems (CPS) \cite{riccobene:abz2014}, but astoundingly the wealth of results on reflection have not been picked up again.

The ups and downs of reflection raise the question of the theoretical foundations of adaptive systems. In this article we make a first step in this direction by means of a {\em behavioural theory} of reflective sequential algorithms. Such a theory is needed to achieve a common understanding of what can be gained by reflection, what the limitations of reflection are. It will show to what extend reflection is supported in various languages. For instance, it is clear that the presence of {\em assert} and {\em retract} in PROLOG provides a limited form of reflection, though it could be argued that this partial support of reflection is sufficient in logic programming.

\subsection{Behavioural Theories}

The ur-instance of a behavioural theory is Gurevich's celebrated {\em sequential ASM thesis} \cite{gurevich:tocl2000}, which states and proves that sequential algorithms are captured by sequential ASMs. A key contribution of this thesis is the language-independent definition of a sequential algorithm by a small set of intuitively understandable postulates on an arbitrary level of abstraction. As the sequential ASM thesis shows, the notion of sequential algorithm includes a form of bounded parallelism, which is a priori defined by the algorithm and does not depend on the actual state. However, parallel algorithms, e.g. for graph inversion or leader election, require unbounded parallelism. A behavioural theory of synchronous parallel algorithms has been first approached by Blass and Gurevich \cite{blass:tocl2003,blass:tocl2008}, but different from the sequential thesis the theory was not accepted, not even by the ASM community despite its inherent proof that ASMs \cite{boerger:2003} capture parallel algorithms. One reason is that the axiomatic definition exploits non-logical concepts such as mailbox, display and ken, whereas the sequential thesis only used logical concepts such as structures and sets of terms. Even the background, that is left implicit in the sequential thesis, only refers to truth values and operations on them.

In \cite{ferrarotti:tcs2016} an alternative behavioural theory of synchronous parallel algorithms (aka ``simplified parallel ASM thesis'') was developed. It was inspired by previous research on a behavioural theory for non-deterministic database transformations \cite{schewe:ac2010}. Largely following the careful motivation in \cite{blass:tocl2003} it was first conjectured in \cite{schewe:abz2012} that it should be sufficient to generalise bounded exploration witnesses to sets of multiset comprehension terms. The rationale behind this conjecture is that in a particular state the multiset comprehension terms give rise to multisets, and selecting one value out each of these multisets defines the proclets used by Blass and Gurevich. The formal proof of the simplified ASM thesis in \cite{ferrarotti:tcs2016} requires among others an investigation in finite model theory. At the same time another behavioural theory of parallel algorithms was developed in \cite{dershowitz:igpl2016}, which is independent from the simplified parallel ASM thesis, but refers also to previous work by Blass and Gurevich. A thorough comparison with the simplified parallel ASM thesis has not yet been conducted.

There have been many attempts to capture asynchronous parallelism as marked in theories of concurrency as well as distribution (see \cite{lynch:1996} for a collection of many distributed or concurrent algorithms). Gurevich's axiomatic definition of partially ordered runs \cite{gurevich:lipari1995} tries to reduce the problem to families of sequential algorithms, but the theory is too strict. As shown in \cite{boerger:ai2016} it is easy to find concurrent algorithms that satisfy sequential consistency \cite{lamport:tc1979}, where runs are not partially ordered. One problem is that the requirements for partially ordered runs always lead to linearisability. The lack of a convincing definition of asynchronous parallel algorithms was overcome by the work on concurrent algorithms in \cite{boerger:ai2016}, in which a concurrent algorithm is defined by a family of agents, each equipped with a sequential algorithm with shared locations. While each individual sequential algorithm in the family is defined by the postulates for sequential algorithms\footnote{A remark in \cite{boerger:ai2016} states that the restriction to sequential algorithms is not really needed. An extension to concurrent algorithms covering families of parallel algorithms is sketched in \cite{schewe:acsw2017}.}, the family as a whole is subject to a concurrency postulate requiring that a successor state of the global state of the concurrent algorithm results from simultaneously applying update sets of finitely many agents that have been built on some previous (not necessarily the latest) states. The theory shows that concurrent algorithms are captured by concurrent ASMs. As in concurrent algorithms, in particular in case of distribution, message passing between agents is more common than shared locations, it has further been shown in \cite{boerger:jucs2017} that message passing can be captured by regarding mailboxes as shared locations, which leads to communicating concurrent ASMs capturing concurrent algorithms with message passing.

\section{Conclusions}\label{sec:schluss}

In this article we investigated a behavioural theory for reflective parallel algorithms (RAs) extending our previous work on reflective sequential algorithms (RSAs) in \cite{schewe:scp2022}. Grounded in related work concerning a behavioural theory for synchronous parallel algorithms \cite{ferrarotti:tcs2016} we developed a set of abstract postulates defining RAs, extended ASMs to reflective abstract state machines (rASMs), and proved that any RA as stipulated by the postulates can be step-by-step simulated by an rASM. The key contributions are the axiomatic definition of RAs and the proof that RAs are captured by rASMs. 

With this behavioural theory we lay the foundations for rigorous development of reflective algorithms and thus adaptive systems. So far the theory covers reflective sequential and parallel algorithms, but not non-deterministic algorithms nor asynchronous concurrent algorithms. So in view of the behavioural theory for concurrent algorithms the next step of the research is to extend also this theory to capture reflection. We envision a part III on reflective concurrent algorithms, which would lay the foundations for the specification of distributed adaptive systems in general. It should be not as difficult as the work in this article, because concurrent algorithms are defined by families of parallel algorithms with concurrent runs \cite{boerger:ai2016}.

Concerning non-determinism, however, there is not yet a behavioural theory available, except for the case of bounded non-determinism in connection with bounded parallelism, which is a simple add-on to the sequential ASM thesis \cite{gurevich:tocl2000,boerger:2003}. Therefore, first such a theory has to be developed before extending it to cover reflection.

Furthermore, for rigorous development extensions to the refinement method for ASMs \cite{boerger:fac2003} and to the logic used for verification \cite{ferrarotti:igpl2017,ferrarotti:amai2018} will be necessary. These will also be addressed in follow-on research.

\section*{CRediT Authorship Contribution Statement}

The authors declare that the submitted work is the result of joint research of both authors. Both Flavio Ferrarotti and Klaus-Dieter Schewe contributed equally to all parts of the submitted article.

\section*{Declaration of Competing Interest}

The authors declare that they have no known competing financial interests or personal relationships that could have appeared to influence the work reported in this paper.

\bibliographystyle{elsarticle-num}
\bibliography{rasm}

\end{document}